\pdfoutput=0

\documentclass[aps,prd,superscriptaddress,nofootinbib,floatfix,preprint]{revtex4-1}

\usepackage{graphicx}
\usepackage{amsmath}
\usepackage{bm}
\usepackage[small,bf]{caption}
\usepackage[subrefformat=parens]{subcaption}
\usepackage{float}
\usepackage{comment}

\newcommand{\beq}{\begin{equation}}
\newcommand{\eeq}{\end{equation}}
\newcommand{\beqa}{\begin{eqnarray}}
\newcommand{\eeqa}{\end{eqnarray}}
\newcommand{\bpr}{\begin{problem}}
\newcommand{\epr}{\end{problem}}
\newcommand{\bcent}{\begin{center}}
\newcommand{\ecent}{\end{center}}
\newcommand{\bfig}{\begin{figure}}
\newcommand{\efig}{\end{figure}}
\newcommand{\bpc}{\begin{picture}}
\newcommand{\epc}{\end{picture}}
\newcommand{\barr}{\begin{array}}
\newcommand{\earr}{\end{array}}
\newcommand{\bitm}{\begin{itemize}}
\newcommand{\eitm}{\end{itemize}}
\newcommand{\bright}{\begin{flushright}}
\newcommand{\eright}{\end{flushright}}
\newcommand{\bminip}{\begin{minipage}}
\newcommand{\eminip}{\end{minipage}}
\newcommand{\btab}{\begin{tabular}}
\newcommand{\etab}{\end{tabular}}

\newcommand{\hiroshima}{Graduate School of Advanced Science and Engineering, Hiroshima University, Kagamiyama, Higashi-Hiroshima, Hiroshima 739-8526, Japan}
\newcommand{\QUP}{International Center for Quantum-field Measurement Systems
for Studies of the Universe and Particles (QUP), KEK, Tsukuba, Ibaraki 305-0801, Japan}

\newcommand{\icr}{Institute for Chemical Research, Kyoto University Uji, Kyoto 611-0011, Japan}
\newcommand{\tokai}{Research Institute of Science and Technology, Tokai University, 4-1-1 Kitakaname, Hiratsuka, Kanagawa 259-1292, Japan}

\newcommand{\kyoto}{Graduate School of Science, Kyoto University, Sakyouku, Kyoto 606-8502, Japan}
\newcommand{\ELINP}{Extreme Light Infrastructure-Nuclear Physics (ELI-NP)/Horia Hulubei National Institute for R\&D in Physics and Nuclear Engineering (IFIN-HH), 30 Reactorului St., P.O. Box MG-6, Bucharest-Magurele, Judetul Ilfov, RO-077125, Romania}
\newcommand{\NILPR}{National Institute for Laser, Plasma and Radiation Physics, 409 Atomistilor  PO Box MG-36, 077125, Magurele, Jud. Ilfov, Romania}

\newcommand{\om}{\omega}
\newcommand{\vth}{\vartheta}

\newcommand{\mcal}[1]{\mathcal{#1}}
\newcommand{\mrm}[1]{\mathrm{#1}}

\newcommand{\Int}[2]{\int_{#1}^{#2}}

\newcommand{\ve}{\varepsilon}

\newcommand{\Tsix}{\mathrm{T}^{6}}

\newcommand{\Equation}[1]{\begin{equation}#1\end{equation}}
\newcommand{\SplitEqn}[1]{\begin{equation}\begin{split}#1\end{split}\end{equation}}
\newcommand{\Align}[1]{\begin{align}#1\end{align}}

\newcommand{\AlignedEqn}[1]{\begin{equation}\begin{aligned}#1\end{aligned}\end{equation}}

\newcommand{\Exp}[1]{\exp\left[#1\right]} 
\newcommand{\ParenB}[1]{\left(#1\right)} 
\newcommand{\CurlyB}[1]{\left\{#1\right\}} 
\newcommand{\BoxB}[1]{\left[#1\right]} 

\newcommand{\vep}{\varepsilon}

\begin{document}
\title{Search for sub-eV axion-like particles 
in a quasi-parallel stimulated resonant photon-photon collider with "coronagraphy"}

\author{Yuri Kirita}\affiliation{\hiroshima}\affiliation{\kyoto}
\author{Airi Kodama}\affiliation{\hiroshima}
\author{Kensuke Homma\footnote{corresponding author}}\affiliation{\hiroshima}\affiliation{\QUP}
\author{Catalin Chiochiu}\affiliation{\ELINP}
\author{Mihai Cuciuc}\affiliation{\ELINP}
\author{Georgiana Giubega}\affiliation{\ELINP}\affiliation{\NILPR}
\author{Takumi Hasada}\affiliation{\hiroshima}
\author{Masaki Hashida}\affiliation{\icr}\affiliation{\tokai}
\author{ShinIchiro Masuno}\affiliation{\icr}
\author{Yoshihide Nakamiya}\affiliation{\ELINP}
\author{Liviu Neagu}\affiliation{\ELINP}\affiliation{\NILPR}
\author{Vanessa Ling Jen Phung}\affiliation{\ELINP}
\author{Madalin-Mihai Rosu}\affiliation{\ELINP}
\author{Shuji Sakabe}\affiliation{\icr}\affiliation{\kyoto}
\author{Stefan Victor Tazlauanu}\affiliation{\ELINP}
\author{Ovidiu Tesileanu}\affiliation{\ELINP}
\author{Shigeki Tokita}\affiliation{\kyoto}
\collaboration{The SAPPHIRES collaboration}

\date{\today}

\begin{abstract}
Axion-like particles (ALPs) have been searched for with a quasi-parallel stimulated
resonant photon-photon collider sensitive to the sub-eV mass range
by focusing two-color near-infrared pulse lasers into a vacuum.
In this work, we have developed a specialized coronagraphy to mitigate the dominant
background photons from optical elements by introducing an eclipse filter.
The observed number of signal-like photons was found to be consistent with
residual background photons from optical elements through an additional test
by degrading the focal point overlapping factor between the two lasers.
We then extended the exclusion region in the relation 
between ALP-photon coupling, $g/M$, and the ALP mass $m$,
reaching the most sensitive point $g/M = 5.45\times10^{-7}\,\mathrm{GeV^{-1}}$ 
at $m = 0.15\,\mathrm{eV}$ for pseudoscalar ALPs. 
\end{abstract}

\maketitle

\section{Introduction}
The strong CP problem is a long standing issue in the sector of quantum chromodynamics (QCD)
in the Standard Model of particle physics.
Peccei-Quinn introduced a new global $U(1)$ symmetry~\cite{Peccei:1977hh,Peccei:1977ur}
to resolve this issue
and appearance of axion is predicted as a kind of pseudo 
Nambu-Goldstone boson~\cite{Weinberg:1977ma,Wilczek:1977pj}
as a result of the $U(1)$ symmetry breaking.
Given the energy scale for the symmetry breaking higher than that of
the electroweak symmetry, the coupling of axion to ordinary matter can be feeble.
Thus axion can be a natural candidate for cold dark matter~
\cite{Preskill:1982cy,Dine:1982ah,Abbott:1982af}.
In addition to axion, more generic axion-like particles (ALPs) are also predicted in various contexts.
The QCD axion model explicitly requires the proportional relation between 
axion mass and axion-matter couplings.
In contrast ALPs do not necessarily require it.
The widely cited mass window for the QCD axion models to account for the dark matter abundance is 
1-100 $\mu$eV. On the other hand, topological defects may cause different predictions on the mass scale.
In particular, a model for string defect in the post-inflationary scenario predicts
lower mass bounds around meV~\cite{Gorghetto}.
Furthermore, an ALP relevant to inflation predicts the mass range up to eV~\cite{MIRACLE} as well. 
So far the QCD axion benchmark models in the mass range 1-100~meV 
indeed have not been fully tested yet. 
Therefore, it is worth testing the sub-eV scale QCD axion and ALP scenarios
as in general as possible.

We address the generic coupling between sub-eV ALPs and two photons
for scalar ($\phi$) and pseudoscalar ($\sigma$) ALPs as follows
\beq
-L_{\phi} = gM^{-1}\frac{1}{4}F_{\mu\nu}F^{\mu\nu}\phi , \hspace{10pt} -L_{\sigma} = gM^{-1}\frac{1}{4}F_{\mu\nu}\tilde{F}^{\mu\nu}\sigma,
\label{eq_phisigma}
\eeq
where dimensionless constant $g$ for a given energy scale $M$ at which a relevant symmetry is broken
is introduced using the electromagnetic field strength tensor
$F^{\mu\nu}=\partial^{\mu}A^{\nu}-\partial^{\nu}A^{\mu}$
and its dual $\tilde{F}^{\mu\nu} \equiv \frac{1}{2}\ve^{\mu\nu\alpha\beta}F_{\alpha\beta}$ with
the Levi-Civita symbol $\ve^{ijkl}$ which is defined as 1 if (i, j, k, l) is an even permutation of (0, 1, 2, 3), 
-1 if it is an odd permutation, and 0 if any index is repeated.

The SAPPHIRES collaboration applied stimulated resonant photon-photon collisions (SRPC)
to the general searches for ALPs~\cite{SAPPHIRES00}, 
which were totally independent of any cosmological and astrophysical assumptions.
In the quasi parallel SRPC, a single pulse laser beam (Ti:Sapphire laser) was focused in the vacuum
and simultaneously another pulse laser beam (Nd:YAG laser) was commonly focused after combing 
the two laser beams along the same optical axis.
The role of the first beam is to create an ALP resonant state via two photon collisions
between stochastically selected two photons within the single focused beam. The role of the second beam
is to induce decay of the produced ALP state into two photons.
When momentum and polarization states of either one of two decayed photons coincide with those
in the inducing beam, the decay rate is enhanced with the proportional dependence 
on the average number of photons in the inducing beam~\cite{SAPPHIRES00}.
As a result of the ALP creation and the stimulated decay, we can expect generation of a signal photon with
a specific energy and direction via energy-momentum conservation between the three laser photons
and the signal one. If we assume photon energies: $\om_c$ in the creation beam and $\om_i$ 
in the inducing beam, the signal photon energy becomes $2\om_c-\om_i$, which is clearly separable
from the beam photon energies.
It can be interpreted as four-wave mixing (FWM) in the vacuum. So this entire scattering process
is also referred to as FWM in the vacuum (vFWM) in the following sections.
Indeed, FWM occurs via atoms through the dipole coupling to photons 
because the kinematical relation between four photons can be $2\om_c \pm \om_i$ in atomic FWM (aFWM).
Therefore, residual atoms in a vacuum chamber and also optical elements to control laser beams
can be natural background sources with respect to vFWM.

Photon-photon scattering is inherently sensitive to the pure quantum nature of the vacuum,
which is absent in the classical Maxwell's equations.
Consequently, numerous proposals have been made to directly probe the vacuum under various
theoretical contexts: for instance,
probing TeV scale quantum gravity~\cite{GammaGammaQG},
probing scalar type dark energy models~\cite{DEPTP,Katsuragawa} and
probing quantum electrodynamics (QED) effects~\cite{GammaGammaQED1, GammaGammaQED2}.
Among these proposals, concrete approaches relevant to QED processes~\cite{Lundstrom, Gies}
have emerged in the literature, particularly with the gradual advancement of high-intensity laser technology. 
Notably, the experimental search photon-photon scattering was conducted earlier in \cite{Bernard}
in the context of QED.
Furthermore, proposals have considered combinations of photon beams with vastly different energy scales, 
such as laser-X-ray beams~\cite{XLaser}, laser-$\gamma$ beams~\cite{GammaLaser} and
all optically produced $\gamma-\gamma$ beams~\cite{GammaGamma}.
Especially in the combination of all optical photon beams, the photon-photon scattering cross section, $\sigma$,
is heavily suppressed, scaling as $\sigma \propto E^6_{cms}$, where $E_{cms}$ is the photon-photon
center of mass system energy. 
Therefore, it is essential to stimulate the spontaneous photon-photon scattering process
by adding an inducing field in order to achieve a sufficiently high interaction rate.
SAPPHIRES's approach shares similarities with the aforementioned proposals in its use of 
an inducing field for stimulation. However, a key distinction lies in the inclusion of 
the s-channel pole and the off-shell contributions in the Breit-Wigner resonance function 
for the ALP production within the uncertainty around the intended $E_{cms}$ energy.
This enhancement mechanism is not expected in the QED-based photon-photon scattering processes, 
as they do not include the exchange of unstable particles such as ALPs.
Thanks to the strong suppression of the QED process in the sub-eV range, 
an ideal pristine environment for the ALP search - essentially free of backgrounds 
from the Standard Model processes - can be realized in stimulated photon-photon scattering 
if the collision point is placed in vacuum~\cite{PTEP2017}. However, in real experimental environments,
atoms in the searching system can be dominant background sources as discussed below.

The SAPPHIRES00 search~\cite{SAPPHIRES00} was conducted with sub-mJ pulse lasers.
We have established the way to evaluate the background yields 
from residual-gas originating aFWM (gas-aFWM), plasma generation, and other ambient noise sources.
In the SAPPHIRES01 search~\cite{SAPPHIRES01},
we have updated the ALP search result with laser energies
that were one order of magnitude higher than those in SAPPHIRES00.
In this search optical-element originating aFWM (opt-aFWM) was 
observed for the first time with the mJ level laser pulses.
We then have introduced the new method for discriminating between
the new background component and signals from vFWM. 
In this paper, we have further increased pulse energies up to the 10~mJ level,
where opt-aFWM became too significant.
Since transmissive optical elements tend to produce a huge amount of opt-aFWM,
we have upgraded the entire optical design by replacing those elements with reflective ones.
In addition we have developed a new method, a kind of coronagraphy, by
introducing an eclipse filter, to significantly suppress the opt-aFWM component.

This paper is organized as follows.
We first introduce classification on possible sources of FWM photons and how to extract vFWM based on coronagraphy.
Second, we explain the upgraded experimental setup used for this search.
Third, we show the measurements and results in this search.
Fourth, a new exclusion region for the relation between ALP coupling with photons and ALP mass is 
provided based on the null result as we will see below.
Finally, the conclusion is given.

\section{Coronagraphy in quasi-parallel SRPC}

\subsection{Possible background sources and extraction of FWM photons}
There are known background processes in this experiment: 
photons from plasma formation at the focal point and atomic four-wave mixing (aFWM) processes.
aFWM is composed of two processes originating from residual gas at the focal point
and originating from optical elements in the path of the combined laser beams.
aFWM photons are produced from the combination of two incident beams
in the same way as signal photons via ALP exchanges, 
whereas plasma-originating photons are produced with just one laser beam focusing.
In order to obtain the number of FWM photons by subtracting plasma-originating photons
from observed photons, searching data are taken by requiring four beam combination patterns
between creation and inducing laser pulse injections, which are define as
\begin{description}
\item[S-pattern] both creation and inducing laser pulses are injected resulting in FWM signals
\item[C-pattern] only creation laser pulses are injected
\item[I-pattern] only inducing laser pulses are injected
\item[P-pattern] both laser pulses are absent, that is, pedestal cases.
\end{description}
The number of FWM photons $n_{\rm{FWM}}$ is estimated by the following relations
\beq\label{nFWM1}
n_{\rm{FWM}} = N_S - N_C - N_I + N_P,
\eeq
where $N_S$, $N_C$, $N_I$, and $N_P$ are the number of observed photons in the individual beam combinations.
Furthermore, since $n_{\rm{FWM}}$ includes the signal photons and background photons, 
$n_{\rm{FWM}}$ is modeled as
\beq\label{nFWM2}
n_{\rm{FWM}} \equiv n_{\rm{vFWM}} + n_{\rm{gas}} + n_{\rm{opt}},
\eeq
where $n_{\rm{vFWM}}$, $n_{\rm{gas}}$, and $n_{\rm{opt}}$ represent the number of 
vacuum originating FWM (vFWM) photons, i.e., signal photons originating from ALP exchanges,
residual-gas originating aFWM (gas-aFWM) photons, 
and optical-element originating aFWM (opt-aFWM) photons, respectively.
vFWM photons can be discriminated from aFWM photons
based on the pressure dependence and the beam cross-section dependence, respectively.
The detailed summary is found in \cite{SAPPHIRES01}.
In the SAPPHIRES01 search, we measured the beam cross-section dependence of the number of observed photons
by changing the beam diameter of the creation laser
while fixing the diameter of the inducing beam.
The number of observed photons increased until the diameter of the creation laser 
reached that of the inducing laser and then saturated after that.
Therefore, we concluded that ALP-originating signal photons, that is, vFWM photons were consistent with null.

\subsection{Coronagraphy with an eclipse filter to mitigate opt-aFWM photons}
\begin{figure}[!hbt]
        \begin{tabular}{cc}
         \centering
                \includegraphics[keepaspectratio, scale=0.8]
                {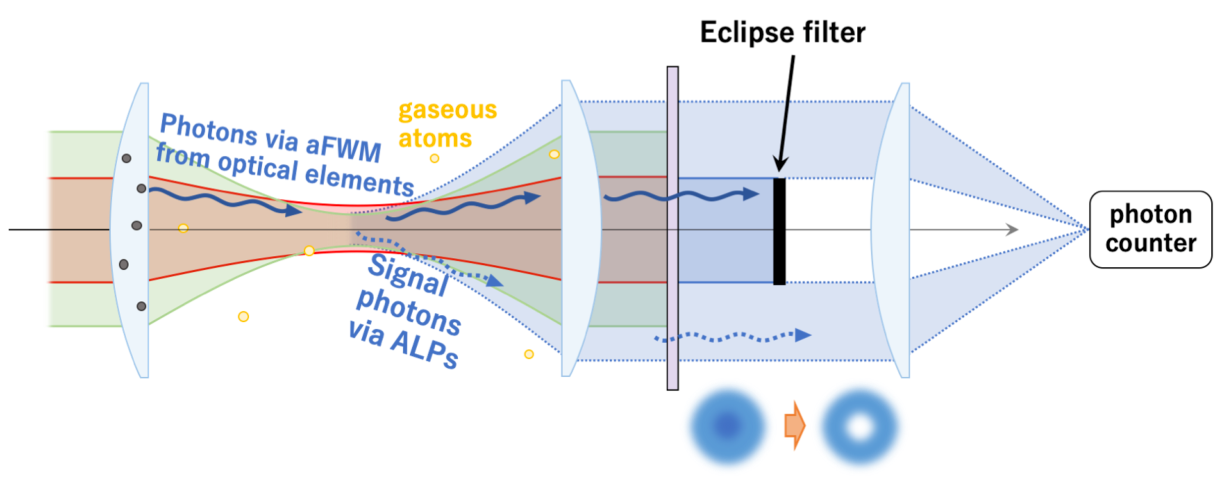}
        \end{tabular}
        \caption{Concept of coronagraphy with an eclipse filter.
        Optical-element originating atomic four-wave mixing (opt-aFWM) photons are generated due to 
        space-time synchronization between the creation and inducing lasers in any optical elements
        such as lenses and mirrors in the system.
        opt-aFWM photons propagate along the path of the inducing laser.
        Therefore, their outgoing angular distribution is the same as that of the inducing laser.
        On the other hand, some fraction of signal photons from the focal point is expected 
        to be emitted to the outside of the angular distribution of the incident beams 
        depending on the exchanged ALP mass~\cite{JHEP2020,SAPPHIRES01}.
        After the second lens, by which the inducing lasers and signal photons can be plane waves
        propagating along the optical axis, an eclipse filter with almost the same diameter 
        as that of the inducing laser is aligned.
        This filter blocks the propagation of opt-aFWM photons.
        Thus signal photons emitted to the peripheral region can be collected and 
        detected by a photon counter with a good signal to noise ratio.}
        \label{Fig1}
\end{figure}

The behavior of the beam cross-section dependence of $n_{\rm{opt}}$ indicates that
they were generated only in the overlap region between two laser beams.
Since two incident beams were coaxially combined,
opt-aFWM photons are kinetically predicted to propagate along the same optical path
and diverge from the focal point as the inducing laser does.
On the other hand, signal photons are expected to be also emitted
outside the diverging cone of the incident beams \cite{JHEP2020}.
We thus can introduce a new method to reduce opt-aFWM photons using
the difference in the outgoing angular distribution
between signal photons and background photons from the interaction point.
The concept of the coronagraphy with an eclipse filter is illustrated in Fig.~\ref{Fig1}.
After collimation by the second lens element, optical-element aFWM photons distribute in the middle region
that approximately corresponds to the propagation of the inducing laser.
The eclipse filter whose diameter is the same as that of inducing laser beam
can block the dominant part of opt-aFWM photons.
The collimated signal photons emitted in the peripheral region
propagate like the solar eclipse and can be detected by a photon counter.

By this coronagraphy, in principle, the detection efficiency to vFWM photons is also diminished.
However, as we discussed in detail in \cite{JHEP2020}, the efficiency reduction is
minor especially in higher ALP mass region to which the quasi-parallel SRPC has a higher sensitivity
thanks to inclusion of lots of combinations from asymmetric photon collisions.
Therefore, even if we introduce the eclipse filter, we can keep a high signal to noise ratio
for the higher mass range.

\section{Experimental Setup and Control}

\begin{figure}[!hbt]
        \begin{tabular}{cc}
         \centering
                \includegraphics[keepaspectratio, scale=0.72]
                {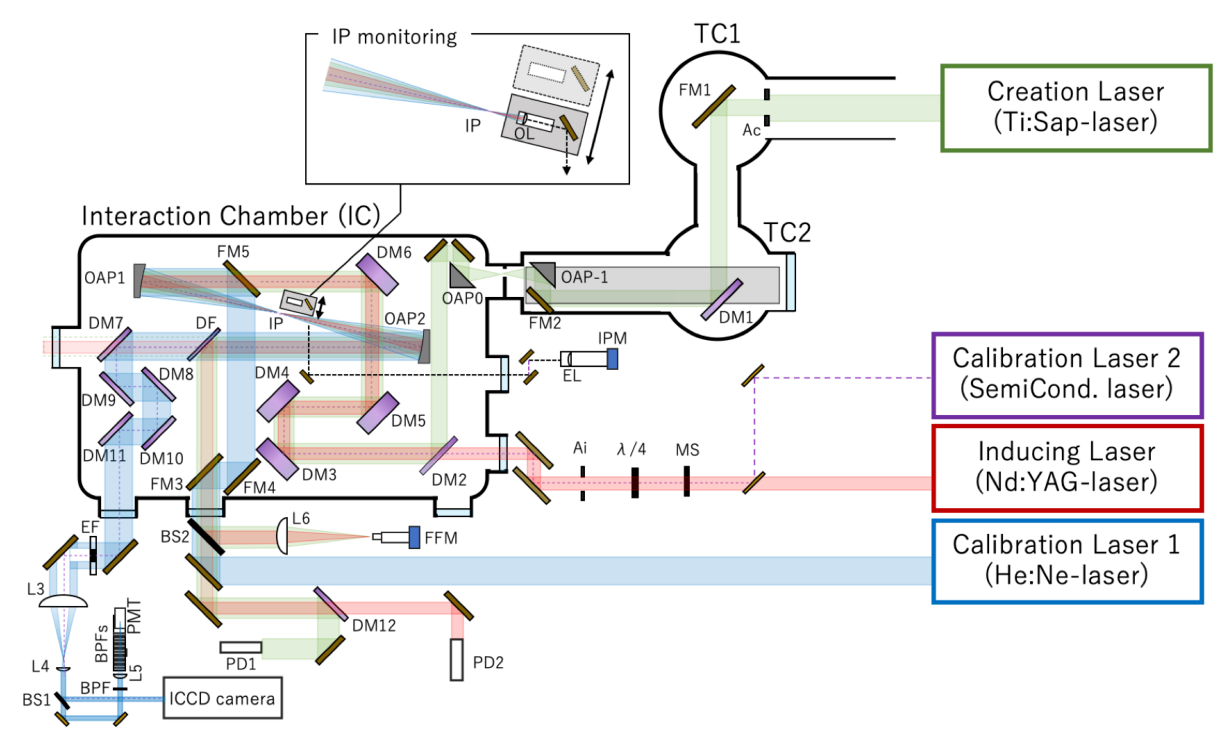}
        \end{tabular}
        \caption{Schematic view of the searching setup.
        This system has been upgraded from the SAPPHIRES01 setup described in \cite{SAPPHIRES01}.
        Ti:Sapphire (816 nm) and Nd:YAG (1064 nm) lasers are used as the creation and 
        the inducing lasers, respectively.
        These lasers are combined at a dichroic mirror (DM2) and
        guided to the off-axis parabolic mirror (OAP1) via dichroic mirrors (DM3-DM6) that
        reflect the creation and inducing lasers but transmit background signal-like photons
        in order to escape them from back before reaching OAP1.
        The incident beams are mostly reflected by a dichroic filter (DF) and
        detected with photo diodes (PD1 and PD2) to produce basic triggers for the waveform recording.
        Signal photons generated at the interaction point (IP) propagate to the outside vacuum chamber 
        through the signal wavelength selections by dichroic mirrors DM7-DM11 and are detected 
        with a photomultiplier tube (PMT) for the photon counting.
        An eclipse filter (EF) blocks the optical-element originating aFWM (opt-aFWM) photons
        before the remaining signal photons reach PMT.
        A semiconductor laser with the wavelength of 450 nm labeled as Calibration laser 2 is combined 
        with the inducing laser and used to place EF at the initial position
        instead of the inducing laser reference, because both 1064 nm and 660 nm are suppressed
        between DM3 and DM11. The hit position by opt-aFWM photons are monitored
        by an image intensifier CCD (ICCD) camera to place the eclipse filter to its final position
        after splitting signal photons by the beam splitter (BS1).
        The calibration laser 1, a He:Ne laser (633 nm), is used with flat mirrors (FM3-FM5)
        to measure the acceptance for signal photons to propagate from IP to PMT.
        This calibration system is removed during the data taking for the search.
        }
        \label{Fig2}
\end{figure}

Figure~\ref{Fig2} shows the schematic view of the searching setup.
This setup was updated from the one used for the SAPPHIRES01 search\cite{SAPPHIRES01}.
We used a Ti:Sapphire laser, the $\rm{T^6}$ system,
available at Institute for Chemical Research in Kyoto University
for the creation laser and a Nd:YAG laser for the inducing laser.
The central wavelength of these lasers were 816 nm and 1064 nm,
and their pulse duration were 40 fs and 9 ns, respectively.
The creation laser was injected through the upstream transport chamber (TC1 and TC2),
where the pressure was maintained at $\sim 10^{-2}$ Pa,
to the interaction chamber (IC) designed for the pressure reaching $10^{-8}$ Pa.
Due to the gap in the pressure values between TC1,2 and IC,
we had introduced a window between the two vacuum systems in the SAPPHIRES01 search~\cite{SAPPHIRES01}.
In this work, however, in order to avoid the self-phase modulation by the optical nonlinear effect
due to the high-intensity laser pulse incidence
and also to mitigate opt-aFWM photons from the window,
we introduced an orifice plate for beam transportation using off-axis parabolic mirrors (OAP-1 and OAP0).
The diameter of the orifice was 10 mm, which was small enough to keep the pressure difference.
Instead, we were forced to implement the new transport system so that the creation laser can
be transported by focusing the field at just after the orifice location.
A linear polarized creation laser (P-polarized) and a circular polarized inducing laser (left-handed)
were combined at dichroic mirror (DM2) in the interaction chamber.

The two beam diameters were selected using irises ($A_c$ and $A_i$).
In SAPPHIRE01~\cite{SAPPHIRES01}, we have sequentially changed the diameter of $A_c$ 
with respect to the fixed diameter of $A_i$ to see the saturation effect of aFWM yield 
at diameters of $A_c$ above that of $A_i$. 
We placed irises instead of reflection type telescopes
because it is better for the diameter to be sequentially variable.
Although we confirmed the similar saturation effect as a function of $A_c$, 
we eventually decided to introduce an eclipse filter to reduce 
the absolute number of opt-aFWM photons,
since it was difficult to sharply determine the unique saturation diameter of $A_c$ due to
increased laser intensities accompanying diffraction components.
The irises were, however, helpful to directly determine the physical beam diameter
even though the small diffraction effect remains, which was indeed small at
IP based on the focal image because the irises were indeed located far from IP.

The central wavelength of FWM photons $\lambda_s$ containing both vFWM and aFWM was
expected to be $\lambda_s = (2\lambda_c^{-1} - \lambda_i^{-1})^{-1} = 660$~nm,
where $\lambda_c$ and $\lambda_i$ are
the central wavelengths of the creation and inducing lasers, respectively.
The wedged thick dichroic mirrors (DM3-DM6) were custom-made to
reflect the creation laser wavelength 816 nm with 98 \% and the inducing laser wavelength 1064 nm with 99 \%
while transmitting the signal wavelength 660 nm with 99 \% to allow signal-like background photons to escape
from back before reaching OAP1.
The coaxially combined incident beams were focused into the interaction point (IP).
The He:Ne laser, named as a calibration laser 1, was guided to OAP1
using flat mirrors (FM3-FM5) which were removed during the search.
OAP2 collected signal photons, and the collimated incident beams were separated from the signal photons
using a thin dichroic filter (DF) and dichroic mirrors (DM7-DM11).
DM7-DM11 were identical custom-made mirrors reflecting 660 nm with greater than 95 \% 
while transmitting around 816 nm with 99 \% and 1064 nm with 95 \% 
to pick up the signal waves among the residual creation and inducing laser beams. 
The DF is a 1 mm thick custom-made dichroic mirror that
reflects 816 nm with 99 \% and 1064 nm with 60 \%,
and transmits 660 nm with 98 \%.

We newly introduced an eclipse filter (EF) outside the interaction chamber to block opt-aFWM photons.
A semiconductor laser with the central wavelength of 450 nm was combined with the inducing laser
and used as the second calibration laser to trace the path of the inducing laser.
To obtain two dimensional information, a 50:50 beam splitter (BS1) was set
and reflected photons were detected with an image-intensifier CCD (ICCD) camera 
(Princeton Instruments, PI-MAX).
This camera has a sensitivity from visible to 900~nm including the central wavelength of the signal photons.
Since the effective image area of ICCD was $12.3 \times 12.3~\rm{mm^2}$,
a reduction optical system with an achromatic lens pair (L3 and L4) was implemented.
The bandpass filters (BPFs) were used to absorb background photons from the residual laser beams.
The signal photons transmitted through BPFs
were detected with a single-photon-sensitive photomultiplier tube (PMT).
In addition to the set of BPFs, a relatively narrow bandpass filter (BPF) was 
added to accept photons in the wavelength range 650-670 nm to reduce plasma-originating photons.
These BPFs were also placed in front of the ICCD camera when ICCD was used.


The time-voltage information from PMT and photodiodes (PDs) were recorded with a waveform digitizer
with a time resolution of 0.5 ns. 
The digitizer was triggered by a basic 10-Hz laser oscillator clock, synchronized to the timing of 
the creation and inducing lasers. The creation laser’s incident rate was reduced to a uniform 5 Hz,
while the inducing laser’s rate was adjusted to a non-uniform 5 Hz 
using a mechanical shutter (MS), generating four staggered trigger patterns 
for offline waveform analysis. 
The four trigger patterns were
(i) two-beam incidence, labeled “S”; 
(ii) inducing-laser-only incidence, labeled “I”; 
(iii) creation-laser-only incidence, labeled “C”; and 
(iv) no-beam incidence, labeled “P”. 
These patterns were cycled sequentially during each data acquisition run, 
ensuring equal shot statistics for each trigger pattern and minimizing systematic uncertainties associated
with subtractions between patterns, as detailed in the following analysis section.

The focused beam profiles were monitored by a CMOS CCD camera (IPM) for both the creation and
inducing beams by transporting the focal images
by an objective lens (OL) properly located from IP combined with an eyepiece lens (EL).
The IP monitoring system was inserted to the focal position in intervals of 
successive searching runs in order to correct the beam drifts between the runs. 
The typical run period was around 30 minutes including the re-adjustment
to minimize the beam drift effects.

\section{Measurements, analysis and result}
\subsection{Spatial distribution of optical-element aFWM photons and the effect of the eclipse filter}
\begin{figure}[!hbt]
        \begin{tabular}{cc}
         \centering
                \includegraphics[keepaspectratio, scale=0.7]
                {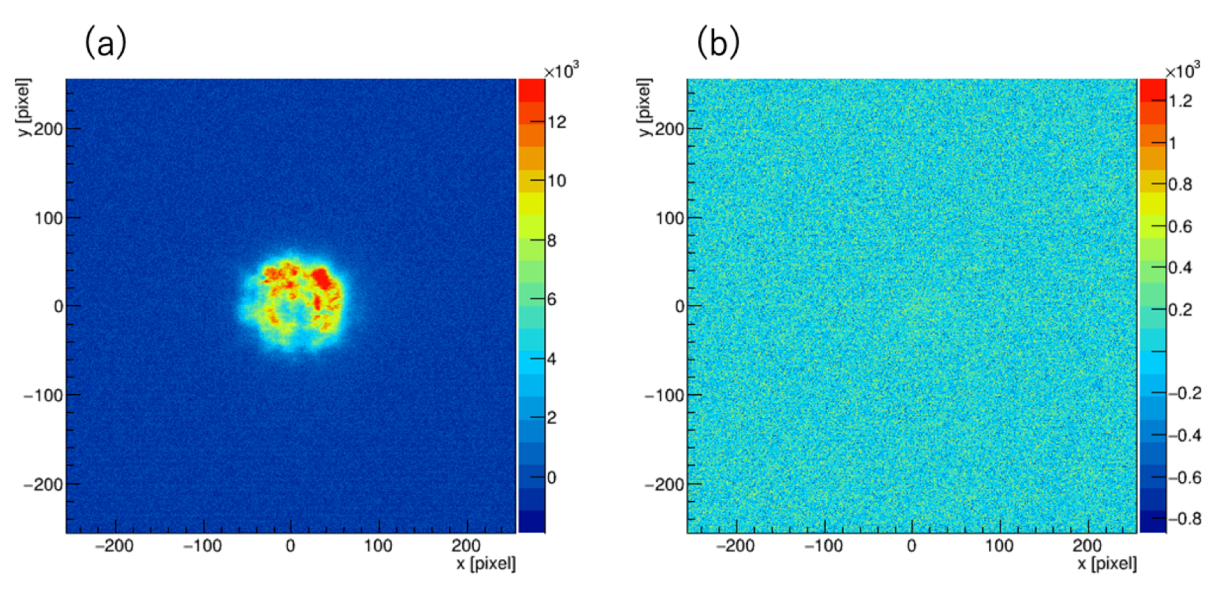}
        \end{tabular}
        \caption{Two dimensional spatial distributions of optical-element originating aFWM (opt-aFWM)
        photons measured by the ICCD camera, which reflects the emission source area of opt-aFWM photons.
        (a) Image of opt-aFWM photons summed over 500 shots of S-pattern without the eclipse filter.
        (b) Image of opt-aFWM photons summed over 500 shots of S-pattern with the aligned eclipse filter.
            Note that the maximum value of the color contour is scaled by 1/10.}
        \label{Fig3}
\end{figure}

Two dimensional distributions of optical-element originating aFWM photons (opt-aFWM)
were measured by the ICCD camera, which reflects the emission source area of opt-aFWM photons.
Figure \ref{Fig3} (a) shows the spatial distribution of opt-aFWM photons 
summed over 500 shots of S-pattern.
The x-y axes represent the camera pixels and the color contour indicates 
normalized ADC-channels of the ICCD camera.
The image is consistent with the shape and size of the inducing laser on the surfaces of
optical elements because the source area of opt-aFWM is limited by the beam cross-section
of the inducing laser.
In order to block the propagation of these opt-aFWM photons,
an eclipse filter was aligned to the inducing laser position using calibration laser 2 in Fig.\ref{Fig2}.
Figure \ref{Fig3} (b) shows the image of opt-aFWM photons 
after setting the eclipse filter consisting of multilayered paper seals on a thin glass plate.
In this experiment, the diameter of the inducing laser was 10 mm,
and the diameter of an eclipse filter was chosen to be 13 mm to reduce the effect from 
diffraction component of the inducing laser.

\subsection{Arrival time distribution in four beam pulse combinations}
To evaluate the number of FWM photons $n_{\rm{FWM}}$ including signal photons using Eq.(\ref{nFWM1}),
we defined four trigger patterns on beam pulse combinations with two incident laser pulses:
when two pulse lasers are incident (Signal pattern, S), 
when only the creation laser is incident (Creation pattern, C),
when only the inducing laser is incident (Inducing pattern, I), 
and when no lasers are incident (Pedestal pattern, P).
Searching data were accumulated by sequentially requiring
sets of the four trigger pattens by setting the mechanical shutter for the inducing laser.
Equation (\ref{nFWM1}) was used to subtract background photons from individual beams, 
likely plasma-originating photons, and pedestal noises.

\begin{figure}[!hbt] 
\begin{tabular}{cc}
 \centering
  \includegraphics[keepaspectratio, scale=0.7]
  {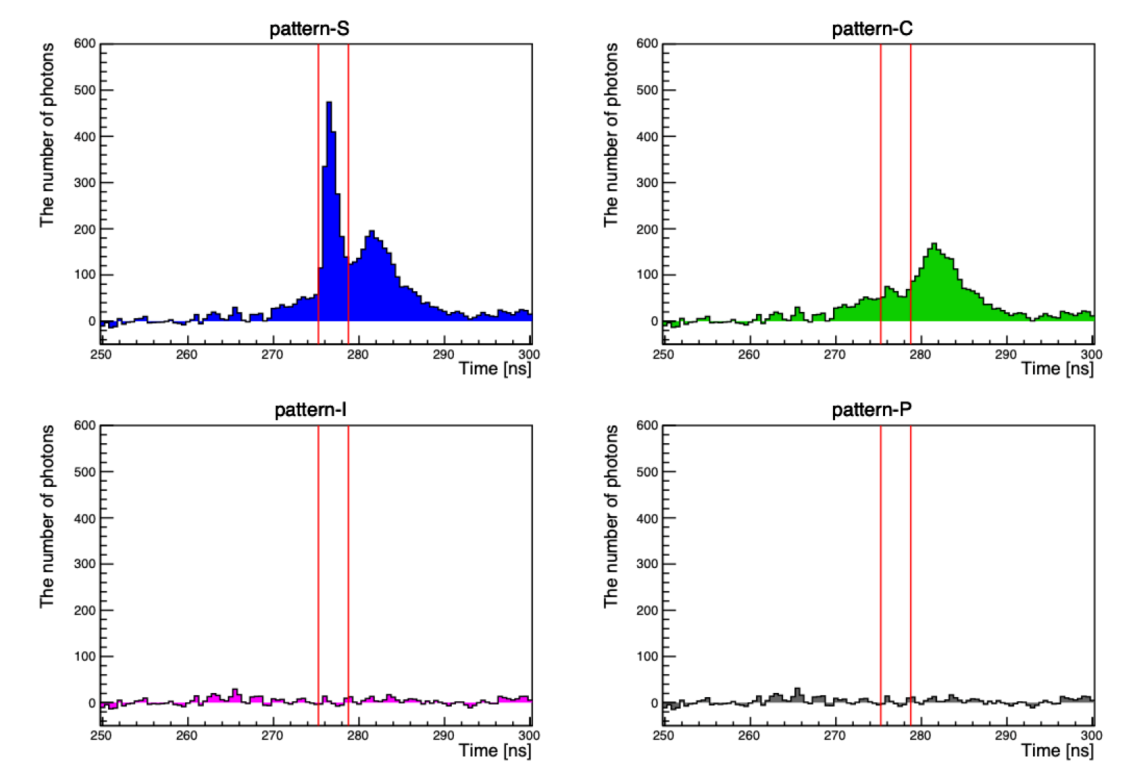}
\end{tabular}
\caption{Arrival time distributions of observed photons
with a single-run shot statistics of 2,500 in individual trigger patterns 
at $4.6 \times 10^{-5}\,\mathrm{Pa}$. 
The red lines represent a time interval within which 
signal-like FWM photons are expected to arrive.
The upper-left, upper-right, lower-left, and lower-right histograms correspond to the 
distributions in S, C, I, and P trigger patterns, respectively.}
\label{Fig4}
\end{figure}

In the previous searches~\cite{SAPPHIRES00,SAPPHIRES01} we have applied the peak finding algorithm
to time vs. voltage waveforms from the waveform digitizer. However, in this data analysis,
we did not apply the peak finder to the waveforms on purpose. This is because of the
higher background level due to increase of the laser intensities.
Applying the finder introduces non-trivial biases when several peaks partially
overlap to each other. Since waveforms can store the falling edge information of analog
signals from PMT, we simply added all waveforms per trigger pattern.
For this analysis the single-photon equivalent charge must be known in advance.
As explained in detail in appendix \ref{Aa} of this paper,
we have obtained the clear linear relation between PMT charges and incident numbers of photons
using a weak pulsed laser by changing the average number of injecting photons per pulse
ranging from 0 to several tens photons. The slope of the line gives the single-photon equivalent charge.
We thus can safely count the number of photons even from charge sums within the signal time window.
Figure \ref{Fig4} shows arrival time distributions of observed photons 
with a single-run shot statistics of 2,500 in individual beam combination patterns 
at $4.6 \times 10^{-5}~\rm{Pa}$, 
where gas-aFWM photons are expected to be completely negligible from 
the known pressure scaling~\cite{SAPPHIRES01}.
The histograms in the upper left, upper right, lower left, and lower right
show the number of observed photons in S, C, I, and P trigger pattern, respectively.
Signal-like FWM photons are expected to arrive within the time interval
subtended by the two red vertical lines.
The timing of the left line was determined from the observed arrival time of the creation laser pluses
at the photodiode (PD1) whose timing resolution was around 40 ps by adding the total time offset 
consisting of the optical path length difference between PD1 and PMT from IP, 
the transition time in PMT and the time difference
for the analogue signals to propagate through the coaxial cables.
The time interval between the two lines was set to 3.5~ns from the calibration data characterizing
the timing structure of analogue signals in the used PMT by taking the dependence of
the incident photon number into account, that is, the slewing effect.
The timing of the right line was determined so that measured time intervals between falling and raising edges
of PMT analogue signals in the calibration data used for obtaining the single-photon equivalent charge
(see Appendix A)
could ensure that the raising edge timings were adequately included in that time interval.
The common threshold voltage value of -2.75 mV for the both edges in the peak finding algorithm,
corresponding to the peak amplitude with the single photon incidence, 
was required to evaluate the time interval when the average number of incident photons per pulse was 
$\lambda = 8.86$ photons in Fig.\ref{Fig12} (d).
This $\lambda$ is indeed consistent with the observed average number of FWM photons 
per shot requring that the number of photons per shot is greater than zero in the searching data.

\begin{figure}[!hbt] 
\begin{tabular}{cc}
 \centering
  \includegraphics[keepaspectratio, scale=0.7]
  {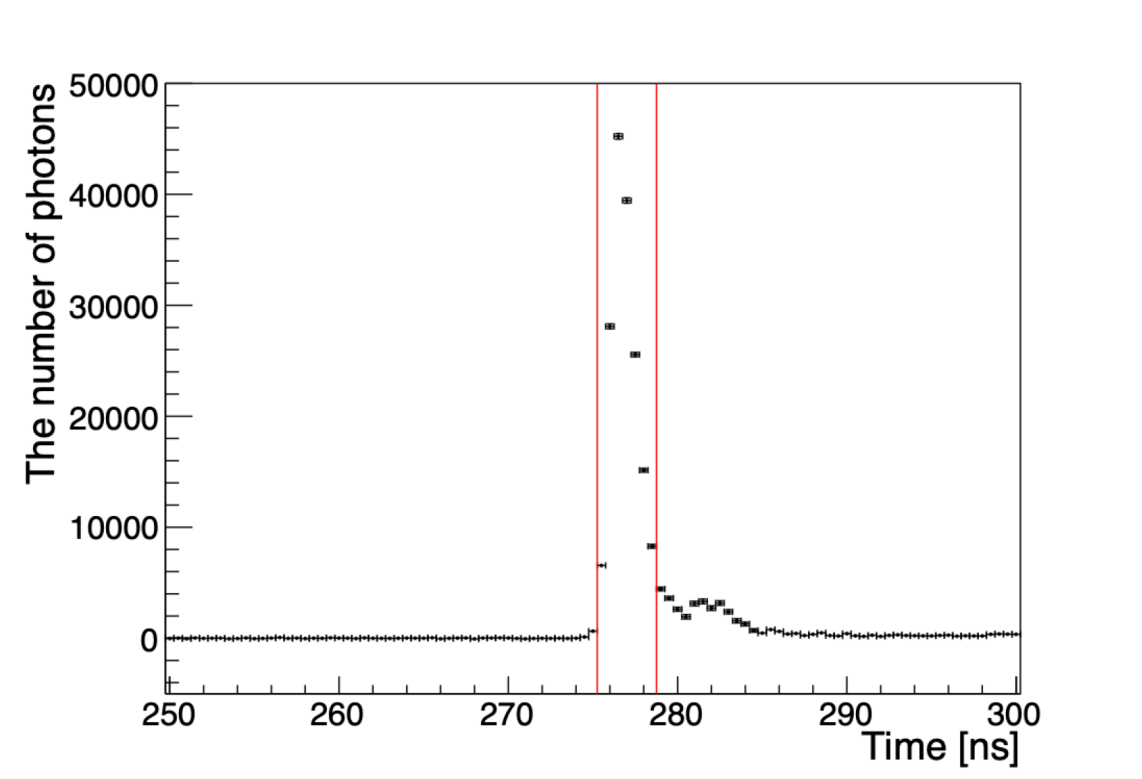}
\end{tabular}
\caption{Arrival time distribution of observed FWM photons after subtraction using Eq.(\ref{nFWM1})
with the total statistics of S-pattern, 22,500 shots.
The red lines represent a time interval within which prompt signal-like FWM photons 
are expected to arrive.}
\label{Fig5}
\end{figure}

Figure \ref{Fig5} shows the arrival time distribution
after subtraction using Eq.(\ref{nFWM1}) with the total statistics of S-pattern, 22,500 shots.
The observed FWM photons are possibly composed of signal photons and residual opt-aFWM photons.
As we discuss in the following section, 
even if the eclipse filter was set at the proper position with the proper size,
a finite amount of opt-aFWM photons could slightly contaminate.
There seems to be a slow background component which
is also seen in the C-pattern histogram in Fig.\ref{Fig4} due to focus 
by only the high-intensity Ti:Sapphire beam.
As background sources contributing to this timing range, we consider the following possibilities:
1) plasma formation, 2) defect effects in SiO${}_2$ included in glass materials such as dichroic mirrors, 
and 3) after-pulse in PMT.
The time structure of the plasma formation is expected to have
the peak at around 70 ps and the typical decay time is known to be around 1 ns~\cite{Plasma}. 
The defects caused by non-bridging oxygen hole centers (NBOHC)
effectively form atom-like level structures, one of which may cause the emission of 650~nm~\cite{NBOHC}.
In addition to the dominant slow decay time of 10-20~$\mu$s, a fast component below 1~$\mu$s was 
also found~\cite{NBOHCfast}, though the fast component itself is still a subject to be investigated.
The complicated level structure possibly may cause the peak-like structure 5-7~ns in the arrival 
time distribution.
In principle, these contributions can be nullified by the proper subtraction between S-pattern and C-pattern.
The after-pulse is an unavoidable feature of the used PMT, which is caused by reflection of electrons
from dynodes back to the photocathode resulting in typically $\sim 5$ ns delayed signals 
associated with individual incident timings. 
This small sub-peak structure should remain as long as
a large number of photons are incident in the proper signal timing window. 
In fact, the after-pulse behavior was seen in the calibration data used to get
the single-photon equivalent charge, even in the case of the single-photon enhanced data set
resulting in the ratio around 0.07.
This ratio is indeed consistent with the peak ratio around 0.05 as seen in Fig.\ref{Fig5}. 
We thus can conclude that the subtraction method without
the bias from the peak finding algorithm properly works for the photon counting.
In any case, these background contributions are dominantly outside the signal timing window
for the scattering process via ALP-exchange which is expected to produce a prompt component.

\subsection{Investigation of the origin of signal-like FWM photons}
From the pressure scaling obtained in the SAPPHIRES01 search~\cite{SAPPHIRES01}, the pressure dependent FWM
component, gas-aFWM, for the used pulse energies at the similar pressure value is 
expected to be too small to explain the observed FWM yield. Thus, there is
a possibility that the observed FWM photons are indeed from the vacuum. However, we need caution.
Although the eclipse filter is effective to suppress the dominant yield from opt-aFWM,
the filter is not perfect because finite diffraction components of the inducing beam 
via apertures in the upstream may cause extension of the inducing field 
beyond the range of the inducing beam diameter on the surfaces of the optical elements.
In order to judge whether the observed FWM photons are from the vacuum or optical elements,
we can utilize a characteristic difference of FWM in the vacuum and in the optical elements
when the two beam overlapping at the focal point is on purpose degraded. The FWM yield from the vacuum
sharply depends on the beam overlapping factor at the focal point as we discuss in the following paragraphs, 
whereas the opt-aFWM is not so sensitive to the focal spot deviation. 
Since the wave vectors in diffraction of the inducing
beam on the surface optical components have a diverging nature, some fraction of the diffraction 
components may always satisfy the phase matching condition to generate atomic FWM 
with the wave vectors of the creation beam on the surface even if the incident directions of the two beams
are slightly different.
Therefore, the beam direction dependence of the residual opt-aFWM is expected to be more gradual.
In the following we have determined the origin of $n_{\rm{FWM}}$ based on this expected feature.

From the particle physics point of view, the vFWM yield corresponds to
the signal yield per pulse collision from stimulated resonant photon-photon scattering
via ALP exchanges, ${\mathcal Y}_{c+i}$, which is factorized as follows~\cite{JHEP2020,SAPPHIRES00}
\beqa\label{eq_Yci}
{\mathcal Y}_{c+i}[1] 
\equiv  \frac{1}{4} N^2_c N_i {\mathcal D}_{qps}\left[s/L^3\right] \overline{\Sigma}_I\left[L^3/s\right].
\mbox{\hspace{2.3cm}}
\eeqa
$N_c$ and $N_i$ are the average numbers of photons in the creation and inducing beams, 
respectively. ${\mathcal D}_{qps}$ is space-time overlapping factor specialized
to the ideal quasi-parallel collision system between the incident beams defined
in Eq.(\ref{Dqps}) in appendix~\ref{Ab} whose full derivation can be found~\cite{GHzMix},
and $\overline{\Sigma}_I$ is the {\it interaction volume rate}, not the
{\it interaction cross section}~\cite{JHEP2020,SAPPHIRES00}, specified in the units
within [\quad], which are expressed in terms of length $L$ and seconds $s$.

Given a set of laser beam parameters $P$,
the number of ALP-originating signal photons,
$n_{\rm{vFWM}}$, as a function of mass $m$ and coupling $g/M$ is expressed as
\beq\label{Nobs}
n_{\rm{vFWM}} = {\cal Y}_{c+i}(m, g/M ; P) t_{a} r \epsilon ,
\eeq
where $t_a$ is a data acquisition time, 
$r$ is a repetition rate of pulsed beams, and
$\epsilon$ is an efficiency of detecting signal photons.
The overall efficiency is defined as $\epsilon \equiv \epsilon_{opt}\epsilon_{det} $,
with $\epsilon_{opt}$ being the optical path acceptance to the signal detector position
and $\epsilon_{det}$  the single photon detection efficiency.
$\epsilon_{opt}$ from IP down to PMT was measured by using calibration 
laser 1 (633 nm) without the narrow 650-670 nm bandpass filter (BPF) 
by measuring the light intensity ratio captured by a CCD camera between the PMT position and IP. 
We then multiplied the vendor's transmittance value of BPF to that acceptance factor in order 
to obtain $\epsilon_{opt}$ from IP to PMT.
$\epsilon_{det}$ was measured using a 532 nm pulse laser in advance of the search. 
We evaluated the absolute detection efficiency by splitting 532 nm pulsed laser beams equally and 
taking the ratio between these energies. The one is measured by a calibrated beam energy meter 
and the other is measured by the PMT with sufficient neutral density filters with measured 
attenuation factors. 
We then corrected the difference of the quantum efficiencies between the 532 nm and around 660 nm lights 
by taking the relative quantum efficiencies provided by the vendor, Hamamatsu Photonics K. K., 
into account.
With respect to a set of $m$ values and an $n_{\rm{vFWM}}$,
a set of coupling $g/M$ can be estimated by numerically solving Eq.(\ref{Nobs}).
The set of laser parameters $P$ is summarized in Table \ref{Tab1}.

\begin{figure}[!hbt] 
\begin{tabular}{cc}
 \centering
  \includegraphics[keepaspectratio, scale=0.7]
  {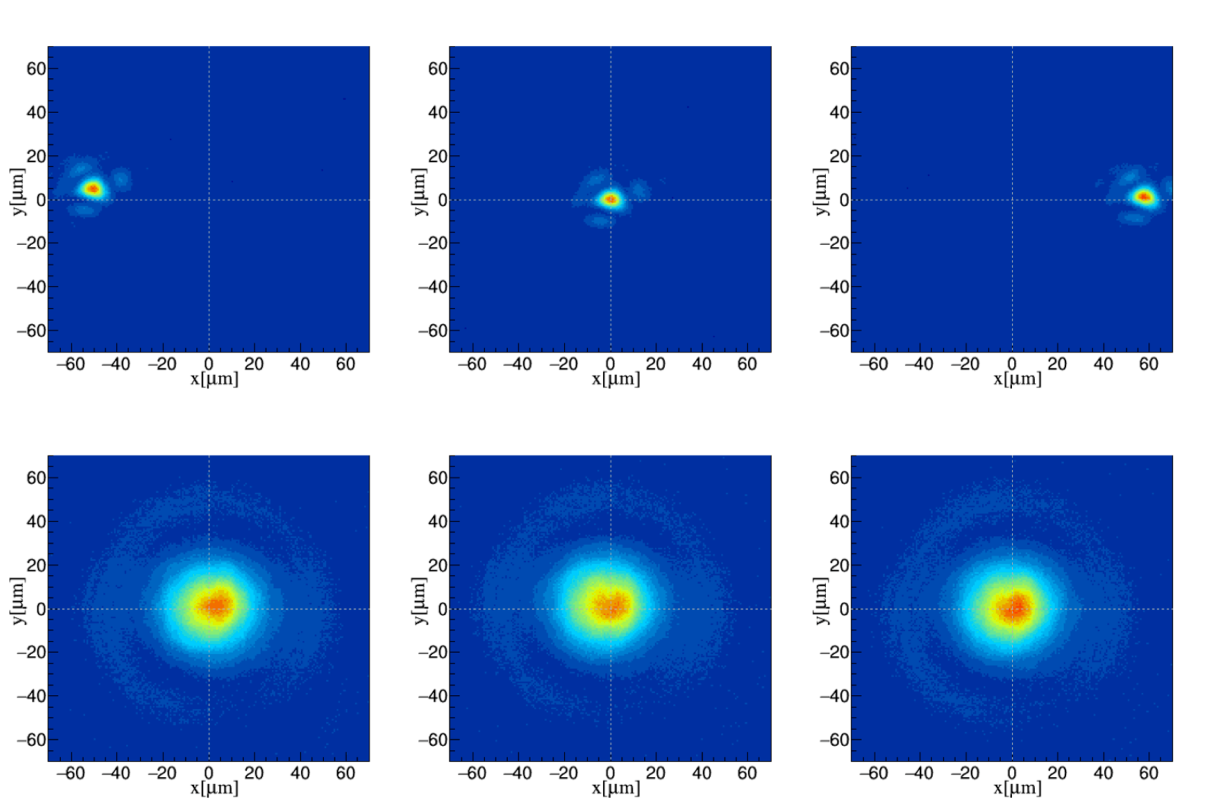}
\end{tabular}
\caption{Degrading degree of overlap between the two beams at IP.
Focal images of the creation (upper row) and inducing beams (lower row) at IP
when the creation beam positions were horizontally shifted by $\pm 2 w_{\rm{YAG}}$
from the best overlapping positions (center figures) with the beam waist~\cite{Yariv}
of the inducing beam, $w_{\rm{YAG}}$.
}
\label{Fig6}
\end{figure}

Figure \ref{Fig6} shows focal images of the creation (upper row) and inducing beams (lower row) at IP
when the creation beam positions were horizontally shifted by $\pm 2 w_{\rm{YAG}}$
from the best overlapping positions (center figures) with the beam waist~\cite{Yariv} 
of the inducing beam, $w_{\rm{YAG}}$.
The beam diameters, twice of beam waists, of the creation and inducing lasers were evaluated 
as 13~$\mu$m and 52~$\mu$m, respectively, by using the IP monitoring system in Fig.\ref{Fig2}.
The focal points were horizontally shifted by using only DM1 in Fig.\ref{Fig2}, that is,
OAP1 and OAP2 were kept fixed.
In addition we further replaced OAP1 and OAP2 with a pair of flat mirrors in
order to limit the source of FWM only to opt-aFWM because there is no focal point
in this case, that is, we can completely exclude the possibility of vFWM.
In order to adjust the pair of the flat mirrors, we used the far field monitor (FFM)
in Fig.\ref{Fig2} so that the two laser directions through the pair of flat mirrors
can be almost identical to the laser directions when DM1 horizontally shifted
positions of the focal points of the creation laser beam at IP.

\begin{figure}[!hbt] 
\begin{tabular}{cc}
 \centering
  \includegraphics[keepaspectratio, scale=0.7]
  {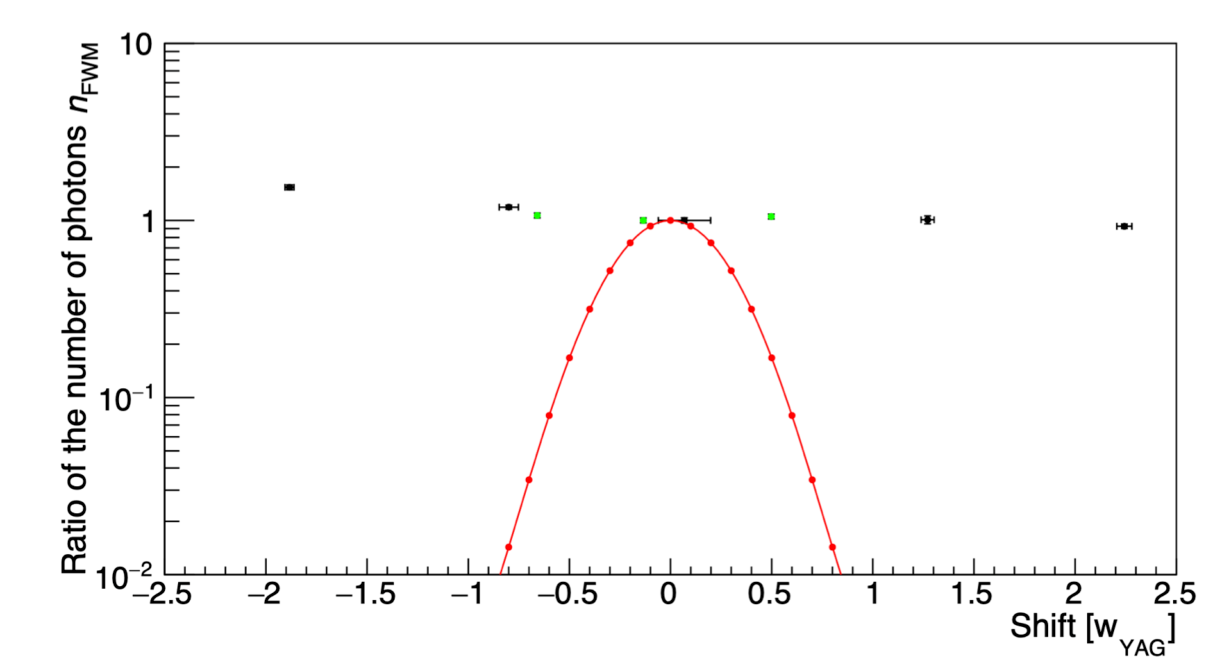}
\end{tabular}
\caption{Comparison of focal-spot overlapping dependences of $n_{\rm{FWM}}$.
Ratios of $n_{\rm{FWM}}$ with respect to $n_{\rm{FWM}}$ at the best overlapping case between
focal points of the creation and inducing beams
are displayed as a function of horizontal focal positions of the creation beam 
in units of beam waist of the inducing laser beam, $w_{\rm{YAG}}$.
Red points: ratios of theoretically expected yields via ALP-exchange with the ALP mass of 0.15 eV
based on the parameters in Tab.\ref{Tab1}.
Black points: ratios of $n_{\rm{FWM}}$ when OAP1 and OAP2 were used for focus and collimation.
Green points: ratios of $n_{\rm{FWM}}$ when the two OAPs were replaced with a pair of flat mirrors,
by which we can guarantee that $n_{\rm{FWM}}$ is composed of only opt-aFWM.
}
\label{Fig7}
\end{figure}

Figure \ref{Fig7} shows the comparison of focal-spot overlapping dependences
between the theoretically expected yield via ALP-exchange
with the ALP mass of 0.15 eV (red points) and observed $n_{\rm{FWM}}$ 
as a function of horizontal focal positions of the creation beam in units of $w_{\rm{YAG}}$. 
The theoretical points are numerically calculated fixing $m$ and $g/M$ in Eq.(\ref{Nobs})
by replacing the idealized ${\mathcal D}_{qps}$ in Eq.(\ref{eq_Yci})
with an experimentally extended space-time overlapping factor
${\mathcal D}_{exp}$~\cite{GHzMix} to implement the spatial drift effects of the focused 
creation beam based on Eq.(\ref{Dexp}) in appendix~\ref{Ab}
with the laser parameters in Tab.\ref{Tab1}.
The black points are $n_{\rm{FWM}}$ when OAP1 and OAP2 were used for focus and collimation, 
whereas green points are $n_{\rm{FWM}}$ when the two OAPs were replaced with a pair of flat mirrors,
by which we can guarantee that $n_{\rm{FWM}}$ is composed of only opt-aFWM.
All the data points are shown as the ratios normalized to the central points around zero in the individual cases
with errors based on quadratic sums between statistical and systematic errors which will be
explained in the next subsection.
Individual data points are derived from single-run measurements, 
making them highly sensitive to the average beam intensity of each run. 
Therefore, in order to compare with the zero shift case, we have corrected the observed number of photons with 
relative weights of $\{E_c(0)/E_c(nw_{\rm{YAG}})\}^2 \{E_i(0)/E_i(nw_{\rm{YAG}})\}$
using measured pulse energies of creation and inducing lasers $E_c$ and $E_i$, respectively,
for runs specified by a $nw_{\rm{YAG}}$ shift with $n \ne 0$.
This correction is based on the known cubic scaling of FWM photons on pulse energies.
The theoretical prediction shows the steep drop even at 0.5 $w_{\rm{YAG}}$.
In contrast, both OAP and flat mirror cases show the similar trend where no such steeply dropping
behavior is seen. The asymmetric tendency in the two cases is understandable from the asymmetry
of OAP geometry with respect to changes of the horizontal incident positions of the creation beam.
Since the flat mirror case can completely exclude the possibility of vFWM from the focal point, 
we conclude that the observed FWM photons in this search is dominated by opt-aFWM.

\subsection{Evaluation of systematic uncertainty}
The dominant sources of systematic errors could consist of
1) shot-by-shot fluctuations of the two laser pulse energies,
2) shot-by-shot fluctuations of temporal overlaps between the two pulsed lasers,
and
3) shot-by-shot focal spot overlaps between the two lasers.
The single run period was determined as $\sim 30$ minutes including re-adjustment of 
the relative focal positions to ensure the effect of 3) was negligibly small.
The drift between focal points of the two lasers was typically 3.3 $\mu$m 
between the start and end of a single run.
The effect of 1) is expected to have the biggest impact because the opt-aFWM yield has 
the cubic dependence on pulse energies. 
The effect of 2) potentially contributes to fluctuations on the background yield too 
though it is difficult to monitor with the present system.
In order to evaluate these combined effects, we have introduced the bootstrap method~\cite{Bootstrap}.
This method can provide a robust test ground to check whether systematic uncertainties are contained or not
even if the underlying systematic correlations are not known {\it a priori} 
by using the observed data set itself.

\begin{figure}[!hbt] 
\begin{tabular}{cc}
 \centering
  \includegraphics[keepaspectratio, scale=0.84]
  {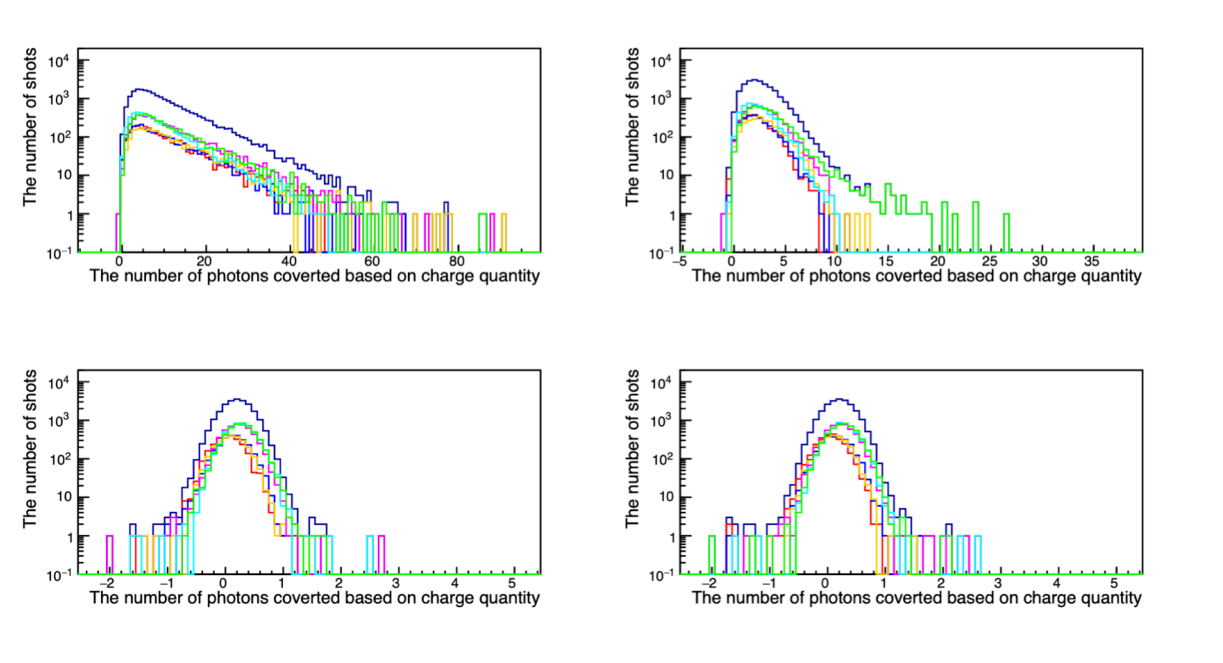}
\end{tabular}
\caption{
The shot-by-shot number of photons converted based on charge quantity
within the signal timing window in the individual trigger patterns: S, C, I and P.
The dark blue histograms represent the distributions for the total sample, 
whereas the colored histograms are classified by individual runs, showing the overall degree of 
systematics included in each run.
}
\label{Fig8}
\end{figure}
\begin{figure}[!hbt] 
\begin{tabular}{cc}
 \centering
  \includegraphics[keepaspectratio, scale=0.84]
  {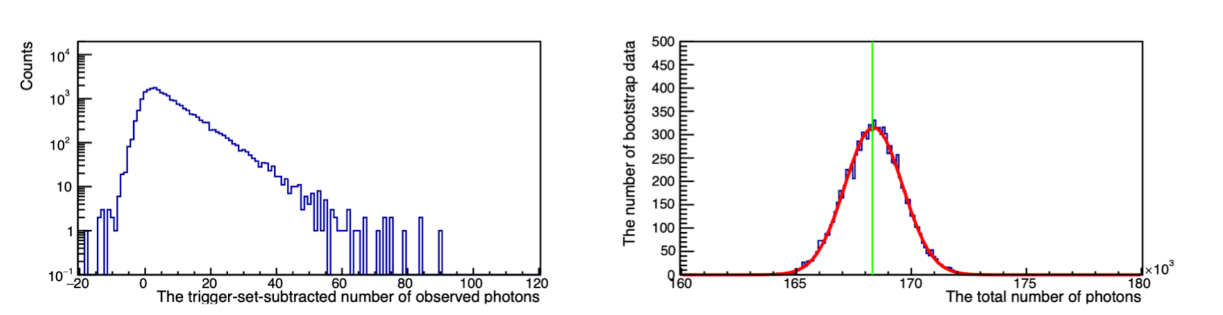}
\end{tabular}
\caption{
Left: The observed number of photons per shot after subtracting within sequential S-C-I-P trigger set. 
Right: The total number of photon distributions in the bootstrap data set of 10000 times
larger statistics with the same number of shot statistics by selecting random numbers
generated according to the actually observed distribution function (left figure) 
which contains real correlations whatever the reasons are.
The vertical line corresponds to the actually observed number of opt-aFWM photons. 
}
\label{Fig9}
\end{figure}

Figure \ref{Fig8} shows the shot-by-shot number of photons converted based on charge quantity
within the signal timing window in the individual trigger patterns: S, C, I and P.
The dark blue histograms represent the distributions for the total sample, 
whereas the colored histograms are classified by individual runs, showing the overall degree of 
systematics included in each run.
%
After subtractions within individual sequential S-C-I-P trigger sets, the number of opt-aFWM photons was
obtained as shown in Fig.\ref{Fig9} (left), which contains potential systematic correlations.
We have created a new bootstrap data set of 10000 times larger statistics 
with the same number of shot statistics by selecting random numbers 
of opt-aFWM photons generated according to the actually observed 
distribution function (left figure) which contains real correlations whatever the reasons are. 
Figure \ref{Fig9} (right) shows the total number of photon distributions
over the 10000 data samples where the vertical line corresponds to the actually observed number 
of opt-aFWM photons. The width of the distribution gives the systematic uncertainty
due to the correlations between 1), 2), 3) and the others.
If the systematic errors were negligibly small, the width of the created distribution must coincide
with the pure statistical uncertainty around the vertical line. 
In fact, the right figure indicates non-statistical contributions from the combined systematic uncertainty.

\subsection{Result}
The analysis indicates that there are no significant vFWM photons in the search data.
The observed background FWM photons are dominated by opt-aFWM.
With the total statistics of 22,500 shots in S-pattern,
we have obtained the result on the number of opt-aFWM photons as
\beqa\label{eq_null}
n_{opt} = 168,308 \pm 546\mathrm{(stat.)} \pm 1,255\mathrm{(syst.)} \quad \mathrm{photons},
\eeqa
where we note that the number of photons were calculated from the measured charge sum
divided by the single-photon-equivalent charge obtained from the independent calibration.
The systematic uncertainty was evaluated by the bootstrap method explained in the above subsection.
We also summarize the set of laser parameters directly relevent to the theoretical calculations
in Tab.\ref{Tab1}.
%

\begin{table}[!ht]
\caption{Experimental parameters directly relevant to the numerical calculations used
for the investigation of the origin of signal-like FWM photons and 
for setting the exclusion limits in the $m - g/M$ parameter spaces.
}
\begin{center}
\begin{tabular}{lr}  \\ \hline
Parameter & Value \\ \hline
%
%
Creation pulse laser & \\
$\quad$ Centeral wavelength $\lambda_c$   & 816 nm\\
$\quad$ Relative linewidth, $\delta\omega_c/<\omega_c>$ &  $1.2\times 10^{-2}$\\
$\quad$ Duration time of pulse, $\tau_{c}$ & 40 fs \\
$\quad$ Measured pulse energy per $\tau_{c}$, $E_{c}$ & 31.4 mJ \\
$\quad$ Pulse energy fraction within 3~$\sigma_{xy}$ focal spot, $f_c$ & 0.6\\
$\quad$ Effective pulse energy per $\tau_c$ within 3~$\sigma_{xy}$ focal spot & $E_{c} f_c = 18.9$~mJ\\
$\quad$ Effective number of creation photons, $N_c$ & $7.8 \times 10^{16}$ photons\\
$\quad$ Beam diameter of pulse, $d_{c}$ & 36.5~mm\\
$\quad$ Polarization & linear (P-polarized state) \\ \hline
Inducing pulse laser & \\
$\quad$ Central wavelength, $\lambda_i$   & 1064~nm\\
$\quad$ Relative linewidth, $\delta\omega_{i}/<\omega_{i}>$ &  $1.0\times 10^{-4}$\\
$\quad$ Duration time of pulse, $\tau_{ibeam}$ & 9~ns\\
$\quad$ Measured pulse energy per $\tau_{ibeam}$, $E_{i}$ & $44.4 $~mJ \\
$\quad$ Linewidth-based duration time of pulse, $\tau_i/2$ & $\hbar/(2\delta\om_{i})=2.8$~ps\\
$\quad$ Pulse energy fraction within 3~$\sigma_{xy}$ focal spot, $f_i$ & 0.8\\
$\quad$ Effective pulse energy per $\tau_i$ within 3~$\sigma_{xy}$ focal spot & $E_{i} (\tau_i/\tau_{ibeam}) f_i = 22.3$~$\mu$J\\
$\quad$ Effective number of inducing photons, $N_i$ & $1.2 \times 10^{14}$ photons\\
$\quad$ Beam diameter of pulse, $d_{i}$ & $10$~mm\\
$\quad$ Polarization & circular (left-handed state) \\ \hline
Focal length of off-axis parabolic mirror, $f$ & 279.1~mm\\
Single-photon detection efficiency, $\epsilon_{det}$ & 1.4 \% \\
Efficiency of optical path from IP to PMT, $\epsilon_{opt}$ & 12 \% \\ \hline
Total number of shots in S-pattern, $W_S$ with $r=5$~Hz   & $22,500$ shots\\
$\delta{n}_{opt}$ & 1369\\
\hline
%
%
\end{tabular}
\end{center}
\label{Tab1}
\end{table}

\section{Exclusion regions in ALP coupling-mass relations}
From the investigation in the previous section, we can conclude that the contribution from $n_{\rm{vFWM}}$ is 
negligibly small and the observed FWM photons are dominated by opt-aFWM.
We thus can calculate the exclusion regions in the coupling-mass relation 
based on the hypothesis that the background distribution is only from the opt-aFWM process. 
Because $n_{opt}$ is obtained from subtractions 
between different trigger patterns whose baseline fluctuations
are expected to individually follow Gaussian distributions,
we can naturally assume that $n_{opt}$ follows the Gaussian distribution as well.

In order to set the upper limit, a confidence level to exclude the null hypothesis that
the observed $n_{\rm{FWM}}$ is all from the opt-aFWM process is defined as
\beq\label{eq_CL}
\mbox{C.L.} =  
\frac{1}{\sqrt{2\pi}\sigma}\int^{\mu+\delta}_{-\infty} e^{-(x-\mu)^2/(2\sigma^2)} dx
= \frac{1}{2} + 
  \frac{1}{\sqrt{2\pi}\sigma}\int^{\mu+\delta}_{0} e^{-(x-\mu)^2/(2\sigma^2)} dx 
\eeq
where $\mu$ is the expected value of an estimator $x$ corresponding to $n_{opt}$ and
$\sigma$ is one standard deviation $\delta n_{opt}$ on the measurement of $n_{opt}$.
We use the acceptance-uncorrected uncertainty for $\delta n_{opt}$
from the quadratic sum of all error components in Eq.(\ref{eq_null}).
For a confidence level of 95 \%, $\mbox{C.L.} = 0.95$ is satisfied
with $\delta = 1.64 \sigma$ based on Eq.(\ref{eq_CL}).
%
In order to evaluate the upper limits on the coupling--mass relation,
we thus numerically solve
\beq
1.64 \delta n_{opt} = {\cal Y}_{c+i}(m, g/M ; P) t_{a} r \epsilon,
\eeq
where $t_a r = W_S = 22,500$ with the repetition rate $r=5$~Hz and
the overall efficiency $\epsilon \equiv \epsilon_{opt}\epsilon_{det}$ in Tab.\ref{Tab1}.

\begin{figure}[H]
	\begin{tabular}{cc}
		\centering
		\includegraphics[keepaspectratio, scale=0.7]{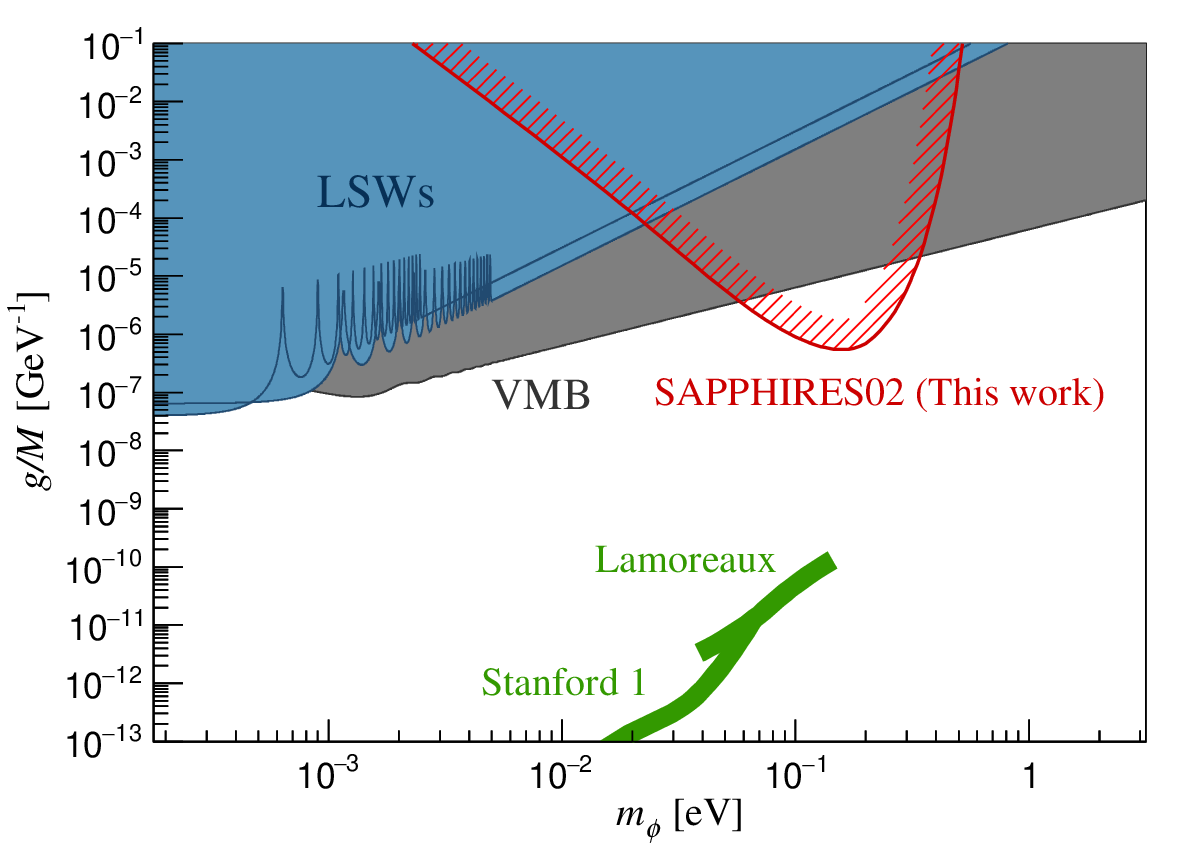}
	\end{tabular}
	\caption{
		Upper limit in the $m_{\phi}-g/M$ parameter space for scalar field exchanges achieved in 
this work, SAPPHIRES02, represented by the red shaded area. 
Limits from the LSW experiments (ALPS \cite{alps} and OSQAR \cite{osqar}) and the VMB experiment (PVLAS \cite{pvlas})  
are shown in the blue and gray areas, respectively. The green lines are exclusion limits from the non-Newtonian force 
searches (Stanford 1 \cite{Stanford1}) and the Casimir force measurement (Lamoreaux \cite{Stanford1}).
	}
	\label{Fig10}
\end{figure}

\begin{figure}[H]
	\begin{tabular}{cc}
		\centering
		\includegraphics[keepaspectratio, scale=0.7]{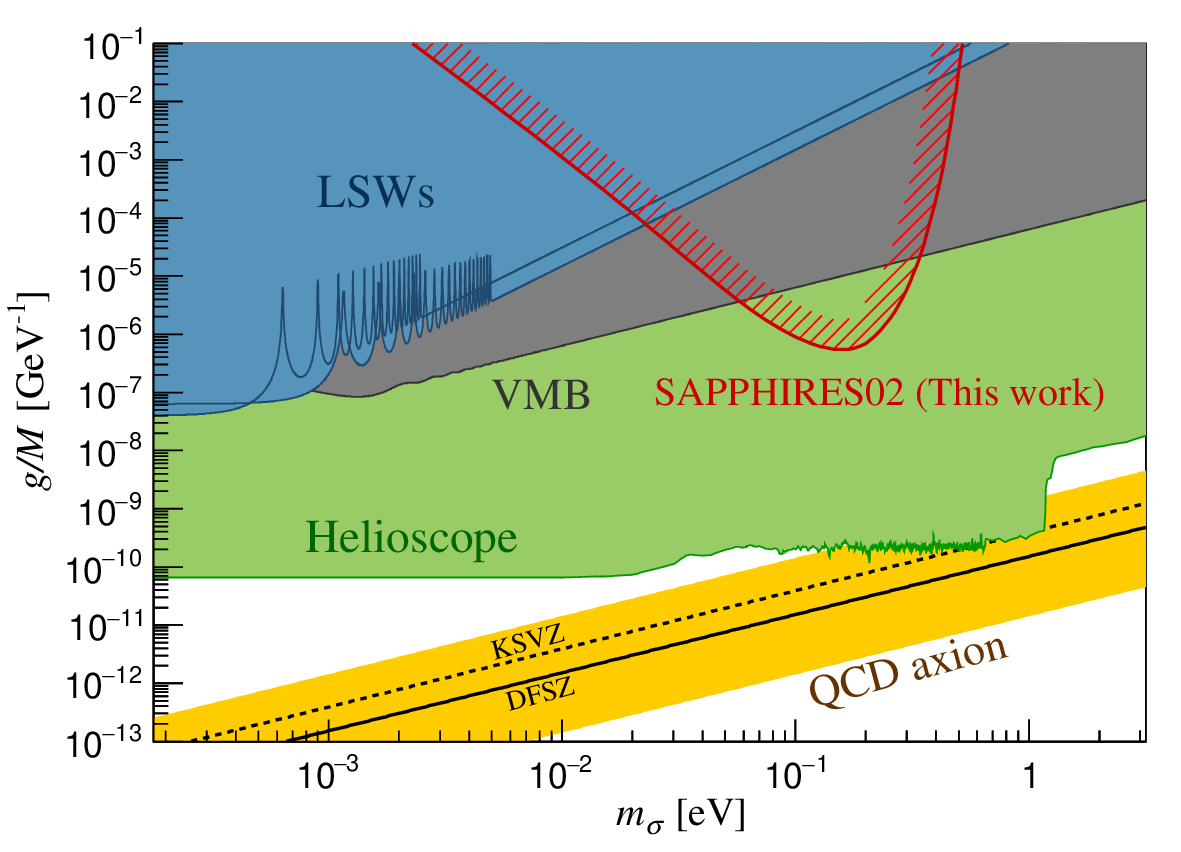}
	\end{tabular}
	\caption{
		Upper limit in the $m_{\sigma}-g/M$ parameter space for pseudoscalar field exchanges achieved 
in this work, SAPPHIRES02, represented by the red shaded area. 
As the cosmological and astrophysical model-independent results,
upper limits from the LSW experiments (ALPS\cite{alps} and OSQAR\cite{osqar})
and the VMB experiment (PVLAS\cite{pvlas}) are shown with the blue and gray areas, respectively. 
The limit from the solar model-dependent helioscope experiment (CAST\cite{cast}) 
is also shown with the green filled areas as a reference. 
The yellow band and the black dashed lines represent QCD axion predictions 
from the KSVZ model \cite{KSVZ} with $0.07 < |E/N - 1.95| < 7$ and $E/N = 0$, respectively.
The solid line is the prediction from the DFSZ \cite{DFSZ} model with $E/N = 8/3$.
	}
	\label{Fig11}
\end{figure}

Figures~\ref{Fig8} and \ref{Fig9} show the updated upper limits
in the coupling--mass relations
for scalar and pseudoscalar fields, respectively, at a 95 \% confidence level
based on the parameter set $P$ in Tab.\ref{Tab1}.
The red shaded areas are from this work.
The ALPS \cite{alps} and OSQAR \cite{osqar} experiments are 
``Light Shining through Wall (LSW)'' experiments
and are filled with blue.
The individual sinusoidal terms in the sensitivities from ALPS and OSQAR are simplified to 1 above $5.0\,\mathrm{meV}$ and $2.5\,\mathrm{meV}$, respectively.
The gray area is the region excluded by the vacuum magnetic birefringence (VMB) 
experiment (PVLAS \cite{pvlas}). In Fig.~\ref{Fig8}, the lines with green are 
excluded by the non-Newtonian force search (Stanford 1 \cite{Stanford1}) and 
the Casimir force measurement (Lamoreaux \cite{Lamoreaux}).
In Fig.~\ref{Fig9}, the green area is the region excluded by the helioscope 
experiment (CAST \cite{cast}). The yellow band is the region 
of the benchmark QCD axion model (KSVZ~\cite{KSVZ}) with $0.07 < |E/N - 1.95| < 7$.
The black dashed and solid lines are predicted
by the KSVZ model with $E/N = 0$ and the DFSZ \cite{DFSZ} model with $E/N = 8/3$, respectively.

\section {Conclusion}
We have searched for sub-eV axion-like particles (ALPs) by
focusing the creation field ($31\,\mathrm{mJ}/40\,\mathrm{fs}$ Ti:Sapphire laser) and
the inducing field ($44\,\mathrm{mJ}/9\,\mathrm{ns}$ Nd:YAG laser)
into the vacuum after combining them along the common optical axis.
Due to the one order of magnitude higher intensity than the previous search~\cite{SAPPHIRES01},
background atomic four-wave mixing (aFWM) became significant as expected from the cubic scaling of FWM
with the laser intensity. In order to mitigate optical-element originating aFWM (opt-aFWM) photons
in the photon counter, we have developed the specialized coronagraphy with the eclipse filter by utilizing
the image intensifier camera.
With respect to the residual signal-like FWM photons, we further have tested
whether those photons originate from residual opt-aFWM or ALP exchanges.
As a result of degrading degree of overlapping between focal spots of the two lasers,
we conclude that opt-aFWM photons dominate the residual FWM photons and no signal FWM is observed.
We then have extended the exclusion region in the relation 
between ALP-photon coupling, $g/M$, and the ALP mass $m$,
reaching the most sensitive point $g/M = 5.45\times10^{-7}\,\mathrm{GeV^{-1}}$ 
at $m = 0.15\,\mathrm{eV}$ for pseudoscalar ALPs. This result is currently the world record in this mass range
among laboratory searches that are completely independent of any cosmological and astrophysical models.

\section*{Acknowledgments}
The $\Tsix$ system was financially supported by the MEXT Quantum Leap Flagship Program (JPMXS0118070187) and the program for advanced research equipment platforms (JPMXS0450300521).

Y. Kirita acknowledges support from the JST, the establishment of university fellowships for the creation of science technology innovation, Grant No. JPMJFS2129, and a Grant-in-Aid for JSPS fellows No. 22J13756 from the Ministry of Education, Culture, Sports, Science and Technology (MEXT) of Japan.

K. Homma acknowledges the support of the Collaborative Research
Program of the Institute for Chemical Research at Kyoto University 
(Grant Nos.\ 2024--95 and 2025-100),
JSPS Core-to-Core Program  (grant number: JPJSCCA20230003),
and Grants-in-Aid for Scientific Research (Nos.\ 21H04474 and 24KK0068)
from the Ministry of Education, Culture, Sports, Science and Technology (MEXT) of Japan.
%
%
 
The authors in ELI-NP acknowledge the support from the Romanian Government and the European Union through the European Regional Development Fund and the Competitiveness Operational Programme (No. 1/07.07.2016, COP, ID 1334). C. Chiochiu, G. Giubega, Y. Nakamiya, L. Neagu, V. L. J. Phung, M. M. Rosu, S. V. Tazlauanu and O. Tesileanu acknowledge the support by Faza 3 (Partea II) a Proiectul Nucleu PN 23 21 01 05. M. Cuciuc acknowledges the support by Faza 4 a Proiectul Nucleu PN 23 21 01 06.

\appendix

\section{Measurement of single-photon equivalent charge}\label{Aa}
In order to obtain the single-photon equivalent charge in the used photomultiplier tube
for the single photon counting,
we have measured charge distributions for a given set of laser pulses with individually
fixed average numbers of incident photons. Although we can fix the individual average numbers
by combining a polarization analyzer pair sandwiching a rotating half-wave plate 
with proper neutral density filters,
we need to determine the absolute average numbers from the following data analysis.
The distributions of charge sum $Q$ within a peak determined by the peak finding algorithm~\cite{SAPPHIRES00}
on the digitized waveforms can be parametrized as the following poissonian-weighted multiple gaussian 
function
\Equation{
	\label{eq:[an-q1]:def_chargedist}
	f_{\lambda:n} (Q) = A \sum_{k = 0}^{n} P_\lambda (k) G_k(Q)
}
with the normalization factor $A$ and the Poissonian distribution $P_\lambda (k)$ defined as
\Equation{
	P_\lambda (k) = \frac{\lambda^k e^{-\lambda}}{k!}
}
where $k$ is the number of incident photons in a laser pulse reflecting fluctuations
around a given average number of incident photons $\lambda$ fixed by the above-mentioned measurement system.
$G_k (Q)$ is the Gaussian distribution including charge fluctuations 
in the process of  multiplication of photoelectrons in the PMT for individual $k$ incident photons, 
which is defined as
\Equation{
	G_k (Q) = \frac{1}{\sqrt{2\pi} \sigma_k} \Exp{-\frac{(Q - \mu_k)^2}{2\sigma_k^2}}
}
where
\Equation{
	\mu_k = \mu_0 + k(\mu_1 - \mu_0) \quad \mbox{and} \ \sigma_k = \sqrt{\sigma_0^2 + k \sigma_1^2}.
}

\begin{figure}[!hbt]
\centering
\includegraphics[keepaspectratio, scale=0.8]{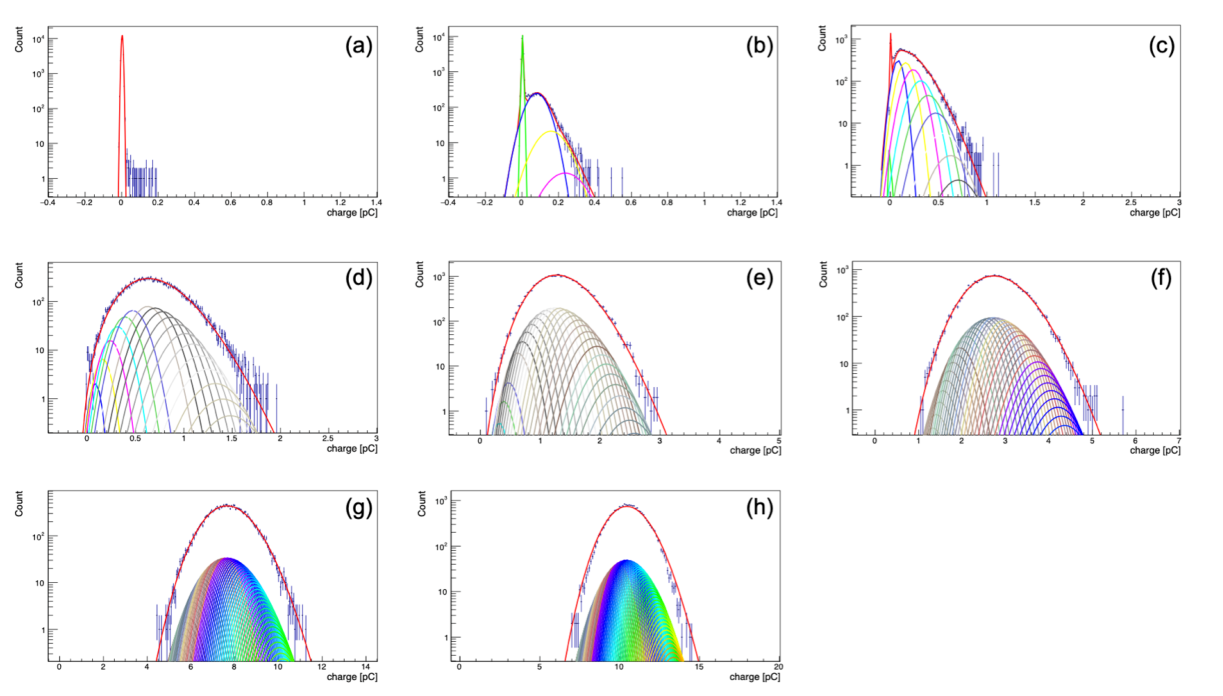}
\caption{
Several-photon-equivalent charge distributions fit with Poissonian-weighted multiple gaussian functions 
$f_{\lambda:n} (Q)$ (red curves).
Poissonian distributions are drawn with various colors.
}
\label{Fig12}
\end{figure}

The fitting procedure is summarized in Fig.\ref{Fig12}.
Firstly $\mu_0$ and $\sigma_0$ can be determined by using zero-photon waveforms without incidence of
laser pulses based on $G_0(Q)$ (Fig.\ref{Fig12}(a)). 
Secondly, $\mu_1$ and $\sigma_1$ can be determined based on $f_{1:3}(Q)$ using the fixed
$\mu_0$ and $\sigma_0$ parameters by using single-photon enriched waveform data 
where the fraction of single photon incident events to zero-photon incident events is 
around 1/5 (Fig.\ref{Fig12}(b)).
Once $\mu_0$, $\sigma_0$, $\mu_1$ and $\sigma_1$ are all fixed by the data sets, 
$f_{\lambda:n}(Q)$ are used to determine $\lambda$ together with $A$ by using waveform data 
with different average numbers of incident photons $\lambda > 1$ (Fig.\ref{Fig12}(c)-(h)).
We note that once $\lambda$ exceeds $\sim 100$, the nonlinear response of PMT starts appearing
as shown in Fig.\ref{Fig12}(h).

\begin{figure}[!hbt]
\centering
\includegraphics[keepaspectratio, scale=0.5]{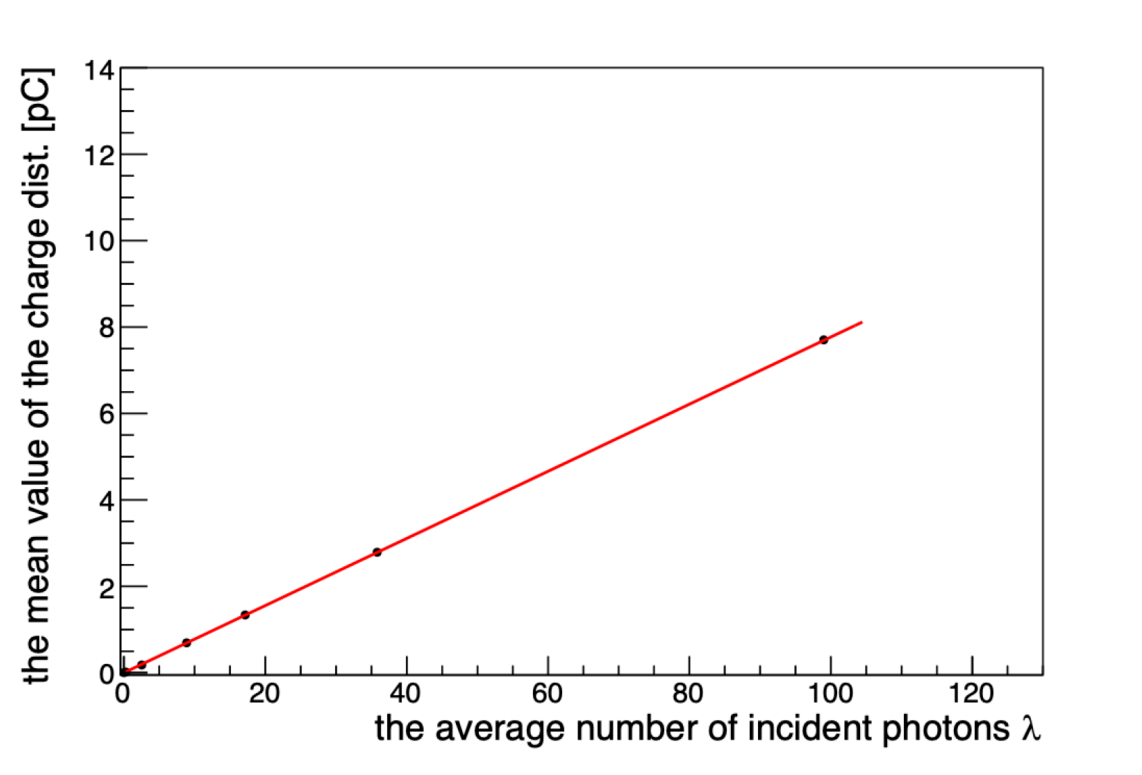}
\caption{The linear relation between the fit average numbers of incident photons $\lambda$ and 
the mean charges in $< 100\,\mrm{photons}$.
These data points were fit with a linear function as the red line
resulting in the slope parameter corresponding to the single-photon equivalent charge.
}
\label{Fig13}
\end{figure}

Figure \ref{Fig13} clearly shows the linear relationship between the fit average numbers of 
incident photons $\lambda$ and the mean charges in the case $< 100\,\mrm{photons}$.
The mean charge is obtained by averaging over charges in photon peaks in the waveforms
belonging to individual $\lambda$ values.
The single-photon equivalent charge is then obtained from the fit slope parameter
\Equation{
	\label{eq:[an-q1]:fitting_afewPhPMT}
	Q_{p.e.} = (7.77 \pm 0.01 \pm 0.08) \times 10^{-2} \, \mrm{[pC/photon]}.
}
We emphasize that the single-photon equivalent charge is robust for the photon counting
in the linear response region even in the case of multiple photon incidence 
with slightly different incident timings, to which the peak finding algorithm is not necessarily 
applicable because it tends to bias the peak shapes due to overlapping of multiple peaks
in the waveform data.

\section{Density overlapping factors, $\mcal{D}_{exp}$ and $\mcal{D}_{qps}$ }\label{Ab}
The most generalized density overlapping factor configured for isolated three pulsed beams
characterizes the degree of spacetime overlap between two creation
pulsed beams $(j = 1, 2)$ and one inducing pulsed beam $(j = 4)$
by assuming that they are focused at a common focal point and 
their peak positions simultaneously arrive at that focal point. 
We thus define the spacetime intersection as the origin of the spacetime coordinates for the pulsed beams.
The full details of the derivation of this factor can be found in appendix of \cite{GHzMix}.

\begin{figure}[!hbt]
        \centering
        \includegraphics[keepaspectratio, scale=0.8]{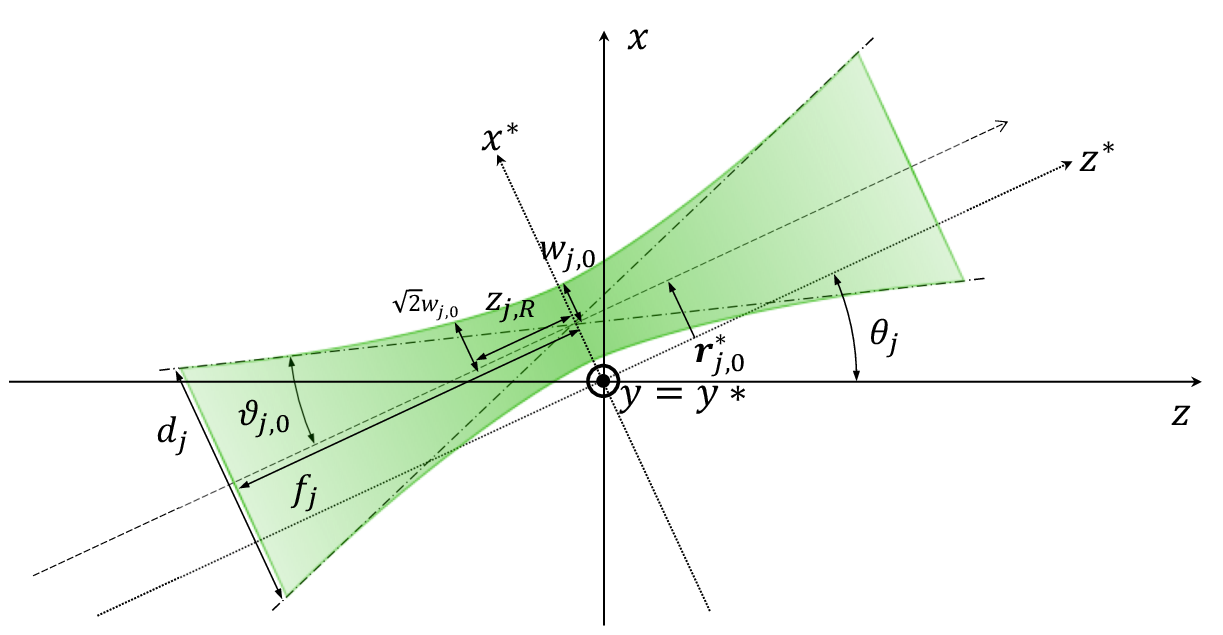}
        \caption{
                Drifted geometry and parameters for a focused laser pulse with incident angle $\theta_j$.
                The laser-beam coordinate system rotated by $\theta_j$ is denoted
                with the asterisk symbol.
                A laser pulse propagates along the dashed arrow parallel to the $z^*$-axis
                with a displacement of $\bm{r}^*_{j,0}$ from that in the $x^*-y^*$ plane.
                The focusing angle between the dashed arrow and the dash-dotted line
                based on geometric optics
                is expressed as $\vth_{j,0} = \tan^{-1} \ParenB{d_j / 2 f_j}$
                using diameter $d_j$ and focal length $f_j$ of beam $j$.
                The beam radius at the focal point is referred to as beam waist $w_{j,0}$,
                and the beam radius along the $z^*$-axis is expressed as $w_j$.
                When $w_j = \sqrt{2} w_{j,0}$, the distance from the focal point
                corresponds to Rayleigh length $z_{j,R}$. This figure and the figure caption
                were extracted from the reference~\cite{GHzMix}.
        }
        \label{Fig14}
\end{figure}

As illustrated in Fig.\ref{Fig14}, we introduce a set of beam parameters which allows deviations
from the idealized beam overlapping case.
This experimentally extended $\cal{D}$-factor is parameterized as
\SplitEqn{ 
        \label{Dexp}
        \mcal{D}_{exp}
                & =
                        \ParenB{\frac{2}{\pi}}^\frac{3}{2}
                        w_{4,0}^2 c_0 \tau_4
                        \Int{- z_{4,\mathit{R}} / c_0}{0} dt
                        \frac{
                                1
                        }{
                                \sqrt{
                                        \ParenB{\sum_j w_j^{-2}}
                                        \ParenB{\sum_j \delta_j}
                                        \ParenB{\sum_j \ParenB{\mu_j^2 + \nu_j^2}}
                                }
                        }
                        \\
                & \qquad
                        \ParenB{
                                \prod_j \frac{1}{w_j^2 c_0 \tau_j}
                        }
                        \Exp{
                                - 2
                                \CurlyB{
                                        \sum_j \ParenB{\xi_j^2 + \eta_j^2 + \zeta_j^2}
                                        -
                                        \frac{
                                                \ParenB{
                                                        \sum_j \ParenB{\mu_j \xi_j + \nu_j \zeta_j}
                                                }^2
                                        }{
                                                \sum_j \ParenB{\mu_j^2 + \nu_j^2}
                                        }
                                }
                        }
}
with the following parameters
\AlignedEqn{
        \label{eq:[th-df]:enum_Df_params}
        \alpha_j
                & =
                        \frac{1}{w_j^2} \cos \theta_j,
        \qquad
        \beta_j
                = \frac{1}{c_0^2 \tau_j^2} \sin \theta_j,
        \qquad
        \delta_j
                = \alpha_j \cos \theta_j + \beta_j \sin \theta_j,
                        \hspace{120pt}
        \\[6pt]
        \mu_j
                & =
                        \frac{1}{\sqrt{\sum_m \delta_m}}
                        \sum_{k,l} \vep_{jkl}
                        \sqrt{\frac{\delta_l}{\delta_k}}
                        \ParenB{
                                \alpha_k \sin \theta_k - \beta_k \cos \theta_k
                        },
        \qquad
        \nu_j
                = \frac{1}{w_j c_0 \tau_j \sqrt{\delta_j}},
        \\[4pt]
        \xi_j
                & =
                        \frac{1}{\sqrt{\sum_m \delta_m}}
                        \sum_{k,l} \vep_{jkl}
                        \sqrt{\frac{\delta_l}{\delta_k}}
                        \ParenB{
                                \alpha_k x^*_{k, 0} + \beta_k c_0 t_k
                        },
        \qquad
        \eta_j
                =       \frac{1}{\sqrt{\sum_m w_m^{-2}}}
                        \sum_{k,l} \vep_{jkl} w_k^{-1} w_l^{-1} y^*_{k,0},
        \\[8pt]
        \zeta_j
                & =
                        \frac{1}{w_j c_0 \tau_j \sqrt{\delta_j}}
                        \ParenB{
                                x^*_{j,0} \sin \theta_j - c_0 t_j \cos \theta_j
                        }.
}
The parameters $\xi_j, \eta_j, \zeta_j$ are caused by the drifts of propagation directions
of individual pulses in the $x^*-y^*$ plane, $x^*_{j, 0}, y^*_{j, 0}$.
Especially, both $\xi_j$ and $\zeta_j$ explicitly include $x^*_{j, 0}$ and $c_0 t_j$
since $x$ and $z$ components are correlated by rotations of laser pulse propagation directions
in the $z-x$ plane.

When the focused spacetime points of individual pulses reach the oring,
the temporal drift becomes $t_{j,0} = 0$ and the spatial drifts
$x^*_{j, 0} = y^*_{j, 0} = 0$ are satisfied.
The parameters for ${\cal{D}}_{exp}$ in such an ideal case without beam drifts are thus summarized as
\AlignedEqn{
        \label{eq:[th-df]:enum_Df_params_qps}
        \alpha_j
                & =
                        \frac{1}{w_j^2} \cos \theta_j,
        \qquad
        \beta_j
                = \frac{1}{c_0^2 \tau_j^2} \sin \theta_j,
        \qquad
        \delta_j
                = \alpha_j \cos \theta_j + \beta_j \sin \theta_j,
                        \hspace{120pt}
        \\[6pt]
        \mu_j
                & =
                        \frac{1}{\sqrt{\sum_m \delta_m}}
                        \sum_{k,l} \vep_{jkl}
                        \sqrt{\frac{\delta_l}{\delta_k}}
                        \ParenB{
                                \alpha_k \sin \theta_k - \beta_k \cos \theta_k
                        },
        \qquad
        \nu_j
                = \frac{1}{w_j c_0 \tau_j \sqrt{\delta_j}},
        \\[4pt]
        \xi_j
                & =
                        \frac{1}{\sqrt{\sum_m \delta_m}}
                        \sum_{k,l} \vep_{jkl}
                        \sqrt{\frac{\delta_l}{\delta_k}}
                        \beta_k c_0 t,
        \qquad
        \eta_j =        0,
        \qquad
        \zeta_j
                =       - \frac{1}{w_j c_0 \tau_j \sqrt{\delta_j}}
                        c_0 t \cos \theta_j.
}

In the case of the quasi-parallel collision system (QPS)
consisting of two beams applied to this paper, 
creation beam ($c$) and inducing beam ($i$),
creation photons are arbitrarily selected from a single common pulse laser
resulting in an ALP creation.
In QPS, since creation and inducing laser pulses co-axially propagate
and are focused by an off-axis parabolic mirror, these incident angles are $\theta_j = 0$.
The corresponding parameters in Eq.\eqref{eq:[th-df]:enum_Df_params} are then as follows
\AlignedEqn{
        \label{eq:[th-df]:enum_QDf_params}
        \alpha_j
                & = \frac{1}{w_j^2},
        \qquad
        \beta_j = 0,
        \qquad
        \delta_j = \frac{1}{w_j^2},
        \qquad
        \mu_j = 0,
        \qquad
        \nu_j = \frac{1}{c_0 \tau_j},
        \hspace{132pt}
        \\[4pt]
        \xi_j
                & =
                        \frac{1}{\sqrt{\sum_m w_m^{-2}}}
                        \sum_{k,l} \vep_{jkl} w_k^{-1} w_l^{-1} x^*_{k, 0},
        \quad
        \eta_j
                =       \frac{1}{\sqrt{\sum_m w_m^{-2}}}
                        \sum_{k,l} \vep_{jkl} w_k^{-1} w_l^{-1} y^*_{k,0},
        \quad
        \zeta_j = - \frac{t_j}{\tau_j}.
}
With the following notations specialized for QPS
$w_c \equiv w_1 = w_2$ and $w_i \equiv w_4$, $\tau_c \equiv \tau_1 = \tau_2$
and $\tau_i \equiv \tau_4$, the $\cal{D}$-factor in QPS is transformed
by substituting the parameters in Eq.\eqref{eq:[th-df]:enum_QDf_params}
into Eq.\eqref{Dexp} as follows
\Align{
        \label{eq:[th-df]:drv_QPSDfactor}
        \mcal{D}_{qps}
                & =
                        \ParenB{\frac{2}{\pi}}^\frac{3}{2}
                        \frac{1}{c_0}
                        \frac{\tau_i}{\tau_c}
                        \frac{1}{\sqrt{\tau_c^2 + 2 \tau_i^2}}
                        w_{i,0}^2
                        \Int{- z_{i,R}/c_0}{0} dt
                        \frac{1}{w_c^2 \ParenB{w_c^2 + 2 w_i^2}}
                        \Exp{
                                - 2 \sum_j \ParenB{\xi_j^2 + \eta_j^2}
                        }.
}
The exponential term in Eq.\eqref{eq:[th-df]:drv_QPSDfactor} expresses
the deviations between the profiles of two laser pulses at the focal point as
\Equation{
        \label{eq:[th-df]:dfm_QDf_devterm}
        \sum_j \ParenB{\xi_j^2 + \eta_j^2}
                =       \frac{2}{w_c^2 + 2 w_i^2}
                        \CurlyB{
                                \ParenB{x^*_{c, 0} - x^*_{i, 0}}^2
                                +
                                \ParenB{y^*_{c, 0} - y^*_{i, 0}}^2
                        }.
}

As the ideal situation in QPS,
we consider the case where time drift $t_{j,0}$ and spatial drift $x^*_{j, 0}, y^*_{j, 0}$ are absent.
Since Eq.\eqref{eq:[th-df]:dfm_QDf_devterm} becomes null, Eq.\eqref{eq:[th-df]:drv_QPSDfactor} is expressed as
\Equation{
        \label{eq:[th-df]:smpl_QPSDfactor}
        \mcal{D}_{qps}
                =       \ParenB{\frac{2}{\pi}}^\frac{3}{2}
                        \frac{1}{c_0}
                        \frac{\tau_i}{\tau_c}
                        \frac{1}{\sqrt{\tau_c^2 + 2 \tau_i^2}}
                        w_{i,0}^2
                        \Int{- z_{i,R}/c_0}{0} dt
                        \frac{1}{w_c^2 (c_0 t) \ParenB{w_c^2 (c_0 t) + 2 w_i^2 (c_0 t)}},
}
where
\Equation{
        \label{eq:[th-df]:dcmp_QDf_fracterm}
        \frac{1}{w_c^2 (c_0 t) \ParenB{w_c^2 (c_0 t) + 2 w_i^2 (c_0 t)}}
                =       \frac{1}{2\ParenB{w_{c,0}^2 \vth_{i,0}^2 -w_{i,0}^2 \vth_{c,0}^2}}
                        \BoxB{
                                \frac{\vth_{c,0}^2}{w_c^2 (c_0 t)}
                                -
                                \frac{\vth_{c,0}^2 + 2\vth_{i,0}^2}{w_c^2 (c_0 t) + 2 w_i^2 (c_0 t)}
                        }.
}
$\mcal{D}_{qps}$ is eventually simplified as
\Equation{
        \label{Dqps}
        \mcal{D}_{qps}
                =       \sqrt{\frac{2}{\pi^3}}
                        \frac{1}{c_0^2}
                        \frac{\tau_i}{\tau_c}
                        \frac{1}{\sqrt{\tau_c^2 + 2 \tau_i^2}}
                        \frac{
                                1
                        }{
                                \vth_{c,0}^2
                                \ParenB{1 - \frac{z_{c,R}^2}{z_{i,R}^2}}
                        }
                        \BoxB{
                                \frac{1}{z_{c,R}}
                                \tan^{-1} \ParenB{\frac{z_{i,R}}{z_{c,R}}}
                                -
                                \frac{1}{Z_{ci}}
                                \tan^{-1} \ParenB{\frac{z_{i,R}}{Z_{ci}}}
                        }
}
with
\Equation{
        \label{eq:[th-df]:def_Zci}
        Z_{ci}
                \equiv
                        \sqrt{
                                \frac{
                                        w_{c,0}^2 + 2 w_{i,0}^2
                                }{
                                        \vth_{c,0}^2 + 2 \vth_{i,0}^2
                                }
                        },
}
which was actually used for the calculation of the upper limits on the coupling for this paper.
This expression exactly coincides with that in the published paper~\cite{SAPPHIRES00},
which is derived starting from the idealized QPS case.


\begin{thebibliography}{99}
\bibitem{Peccei:1977hh}
R.~D.~Peccei and H.~R.~Quinn,
``CP Conservation in the Presence of Instantons'',
Phys. Rev. Lett. \textbf{38}, 1440-1443 (1977)
doi:10.1103/PhysRevLett.38.1440.

\bibitem{Peccei:1977ur}
R.~D.~Peccei and H.~R.~Quinn,
``Constraints Imposed by CP Conservation in the Presence of Instantons'',
Phys. Rev. D \textbf{16}, 1791-1797 (1977)
doi:10.1103/PhysRevD.16.1791.

\bibitem{Weinberg:1977ma}
S.~Weinberg,
``A New Light Boson?'',
Phys. Rev. Lett. \textbf{40}, 223-226 (1978)
doi:10.1103/PhysRevLett.40.223.

\bibitem{Wilczek:1977pj}
F.~Wilczek,
``Problem of Strong  $P$  and  $T$  Invariance in the Presence of Instantons'',
Phys. Rev. Lett. \textbf{40}, 279-282 (1978)
doi:10.1103/PhysRevLett.40.279.

\bibitem{Preskill:1982cy}
J.~Preskill, M.~B.~Wise and F.~Wilczek,
``Cosmology of the Invisible Axion'',
Phys. Lett. B \textbf{120}, 127-132 (1983)
doi:10.1016/0370-2693(83)90637-8.

\bibitem{Dine:1982ah}
M.~Dine and W.~Fischler,
``The Not So Harmless Axion'',
Phys. Lett. B \textbf{120}, 137-141 (1983)
doi:10.1016/0370-2693(83)90639-1.

\bibitem{Abbott:1982af}
L.~F.~Abbott and P.~Sikivie,
Phys. Lett. B \textbf{120}, 133-136 (1983)
doi:10.1016/0370-2693(83)90638-X.

\bibitem{Gorghetto}
M.~Gorghetto, E.~Hardy and G.~Villadoro,
SciPost Phys. \textbf{10}, no.2, 050 (2021)
doi:10.21468/SciPostPhys.10.2.050
[arXiv:2007.04990 [hep-ph]].

\bibitem{MIRACLE}
Ryuji Daido et al., "The ALP miracle revisited", Journal of High Energy Physics, 02 (2018) 104, 16th February 2018.


\bibitem{SAPPHIRES00}
K.~Homma \textit{et al.} [SAPPHIRES],
JHEP \textbf{12}, 108 (2021)
doi:10.1007/JHEP12(2021)108
[arXiv:2105.01224 [hep-ex]].

\bibitem{GammaGammaQG}
H.~Davoudiasl,
Phys. Rev. D \textbf{60}, 084022 (1999)
doi:10.1103/PhysRevD.60.084022
[arXiv:hep-ph/9904425 [hep-ph]].

\bibitem{DEPTP}
Y.~Fujii and K.~Homma,
Prog. Theor. Phys. \textbf{126}, 531-553 (2011)
[erratum: PTEP \textbf{2014}, 089203 (2014)]
doi:10.1143/PTP.126.531
[arXiv:1006.1762 [gr-qc]].


\bibitem{Katsuragawa}
T.~Katsuragawa, S.~Matsuzaki and K.~Homma,
Phys. Rev. D \textbf{106}, no.4, 044011 (2022)
doi:10.1103/PhysRevD.106.044011
[arXiv:2107.00478 [gr-qc]].

\bibitem{GammaGammaQED1}
B.~De Tollis,
Nuovo Cim. \textbf{32}, no.3, 757-768 (1964)
doi:10.1007/BF02735895

\bibitem{GammaGammaQED2}
B.~De Tollis,
Nuovo Cim. \textbf{35}, no.4, 1182-1193 (1965)
doi:10.1007/BF02735534

\bibitem{Gies}
H.~Gies, F.~Karbstein, C.~Kohlf\"urst and N.~Seegert,
Phys. Rev. D \textbf{97}, no.7, 076002 (2018)
doi:10.1103/PhysRevD.97.076002
[arXiv:1712.06450 [hep-ph]].

\bibitem{Lundstrom}
E.~Lundstrom, G.~Brodin, J.~Lundin, M.~Marklund, R.~Bingham, J.~Collier, J.~T.~Mendonca and P.~Norreys,
Phys. Rev. Lett. \textbf{96}, 083602 (2006)
doi:10.1103/PhysRevLett.96.083602
[arXiv:hep-ph/0510076 [hep-ph]].

\bibitem{Bernard}
D.~Bernard, F.~Moulin, F.~Amiranoff, A.~Braun, J.~P.~Chambaret, G.~Darpentigny, G.~Grillon, S.~Ranc and F.~Perrone,
Eur. Phys. J. D \textbf{10}, 141 (2000)
doi:10.1007/s100530050535
[arXiv:1007.0104 [physics.optics]].

\bibitem{XLaser}
F.~Karbstein and E.~A.~Mosman,
Phys. Rev. D \textbf{100}, no.3, 033002 (2019)
doi:10.1103/PhysRevD.100.033002
[arXiv:1906.10122 [physics.optics]].

\bibitem{GammaLaser}
Y.~Nakamiya and K.~Homma,
``Probing vacuum birefringence under a high-intensity laser field with gamma-ray polarimetry at the GeV scale,''
Phys. Rev. D \textbf{96}, no.5, 053002 (2017)
doi:10.1103/PhysRevD.96.053002
[arXiv:1512.00636 [hep-ph]].

\bibitem{GammaGamma}
K.~Homma, K.~Matsuura and K.~Nakajima,
PTEP \textbf{2016}, no.1, 013C01 (2016)
doi:10.1093/ptep/ptv176
[arXiv:1505.03630 [hep-ex]].

\bibitem{PTEP2017}
K.~Homma and Y.~Toyota,
PTEP \textbf{2017}, no.6, 063C01 (2017)
doi:10.1093/ptep/ptx069
[arXiv:1701.04282 [hep-ph]].

\bibitem{SAPPHIRES01}
Y.~Kirita \textit{et al.} [SAPPHIRES],
``Search for sub-eV axion-like particles in a stimulated resonant photon-photon collider with two laser beams based on a novel method to discriminate pressure-independent components,''
JHEP \textbf{10}, 176 (2022)
doi:10.1007/JHEP10(2022)176
[arXiv:2208.09880 [hep-ex]].

\bibitem{Plasma}
A. E. Martirosyan, C. Altucci, A. Bruno, C. de Lisio, A. Porzio, S. Solimeno, 
``Time evolution of plasma afterglow produced by femtosecond laser pulses,''
J. Appl. Phys. 15 November 2004; 96 (10) 5450 – 5455

\bibitem{NBOHC}
Arnaud Zoubir, Martin Richardson, Lionel Canioni, Arnaud Brocas, and Laurent Sarger, 
``Optical properties of infrared femtosecond laser-modified fused silica and application to waveguide fabrication,''
J. Opt. Soc. Am. B 22, 2138-2143 (2005)

\bibitem{NBOHCfast}
L. Skuja, N. Ollier, K. Kajihara, Lu,
``minescence of non-bridging oxygen hole centers as a marker of particle irradiation of $\alpha$-quartz,''
Radiation Measurements, 135 (2020) 106373

\bibitem{JHEP2020}
K.~Homma and Y.~Kirita,
``Stimulated radar collider for probing gravitationally weak coupling pseudo Nambu-Goldstone bosons,''
JHEP \textbf{09}, 095 (2020)
doi:10.1007/JHEP09(2020)095
[arXiv:1909.00983 [hep-ex]].

\bibitem{Yariv}
Amnon Yariv, {\it Optical Electronics in Modern Communications} Oxford University Press (1997).

\bibitem{Bootstrap}
B. Efron, 
``Bootstrap Methods: Another Look at the Jackknife.,'' 
Ann. Statist. 7 (1) 1 - 26, January, 1979. 

\bibitem{GHzMix}
K.~Homma, Y.~Kirita, T.~Miyamaru, T.~Hasada and A.~Kodama,
Phys. Rev. D \textbf{110}, no.9, 092017 (2024)
doi:10.1103/PhysRevD.110.092017
[arXiv:2405.03577 [hep-ph]].


\bibitem{alps}
K. Ehret et al. (ALPS), Phys. Lett. B {\bf 689}, 149 (2010).

\bibitem{osqar}
R. Ballou {\it et al.} (OSQAR), Phys. Rev. D {\bf92}, 9, 092002 (2015).

\bibitem{pvlas}
A. Ejlli {\it et al.}, Physics Reports 871 (2020) 1–74.

\bibitem{Stanford1}
J. Chiaverini {\it et al.}, Phys. Rev. Lett. {\bf 90}, 151101 (2003).

\bibitem{Lamoreaux}
S. K. Lamoreaux, Phys. Rev. Lett. {\bf 78}, 5 (1997);  {\bf 81}, 5475 (1998).

\bibitem{cast}
E. Arik {\it et al.} (CAST), J. Cosmol. Astropart. Phys. {\bf 02}, 008 (2009);
M. Arik {\it et al.} (CAST), Phys. Rev. Lett. {\bf 107}, 261302 (2011);
M. Arik {\it et al.} (CAST), Phys. Rev. Lett. {\bf112}, 9, 091302 (2014);
V. Anastassopoulos {\it et al.} (CAST), Nature Phys. {\bf13}, 584 (2017).

\bibitem{KSVZ}
J. E. Kim, Phys. Rev. Lett. 43, 103 (1979);
M. Shifman, A. Vainshtein, and V. Zakharov, Nucl. Phys. B166, 493 (1980).

\bibitem{DFSZ}
M. Dine, W. Fischler, and M. Srednicki, Phys. Lett. 104B, 199 (1981);
A. Zhitnitskii, Sov. J. Nucl. Phys. 31, 260 (1980).

\end{thebibliography}
\end{document}